\documentclass[twocolumn,epjc3]{svjour3mod}
\usepackage{hyperref}
\usepackage{graphicx,colordvi}
\usepackage{amsmath,amssymb,slashed,cite}
\usepackage{booktabs,tabulary}
\usepackage{dsfont}
\usepackage{color}
\usepackage{epstopdf}  
\usepackage{fix-cm}
\makeatletter
\newcommand*{\textoverline}[1]{$\overline{\hbox{#1}}\m@th$}
\makeatother

\newcommand{\be}{\begin{equation}}
\newcommand{\ee}{\end{equation}}
\newcommand{\bea}{\begin{eqnarray}}
\newcommand{\eea}{\end{eqnarray}}
\newcommand{\no}{\notag \\}

\newcommand{\indR}{\text{\scriptsize R}}

\newcommand{\fp}{\text{fp}}

\newcommand{\ubar}{\bar u}
\newcommand{\dbar}{\bar d}

\newcommand{\fbar}{\bar f}

\newcommand{\QCD}{\mbox{\tiny Q\hspace{-0.05em}CD}}
\newcommand{\QED}{\mbox{\tiny QED}}

\newcommand{\MeV}{\,\text{MeV}}
\newcommand{\GeV}{\,\text{GeV}}
\renewcommand{\Im}{\text{Im}\,}
\newcommand{\inel}{\text{inel}}
\newcommand{\el}{\text{el}}
\newcommand{\disp}{\text{disp}}

\newcommand{\as}{\text{as}}

\newcommand{\sumf}{\mbox{$\sum_f$}}
\newcommand{\nuth}{\nu_\text{\tiny th}}
\newcommand{\xth}{x_\text{\tiny th}}
\newcommand{\alphaem}{\alpha_\text{\tiny em}}

\newcommand{\al}{&\!}

\newcommand{\bsp}{\begin{sloppypar}}
\newcommand{\esp}{\end{sloppypar}}

\RequirePackage{graphicx}

\journalname{Eur. Phys. J. C}
\begin{document}

\title{Sum rule for the Compton amplitude and implications\\ for the proton-neutron mass difference}
\author{J.\ Gasser\thanksref{addr1}
        \and
        H.\ Leutwyler\thanksref{addr1}
        \and
        A.\ Rusetsky\thanksref{addr2} 
}
\institute{Albert Einstein Center for Fundamental Physics, Institut f\"ur theoretische Physik, Universit\"at Bern, Sidlerstrasse 5, CH--3012 Bern, Switzerland\label{addr1}
\and
Helmholtz-Institut f\"ur Strahlen- und Kernphysik (Theorie) and Bethe Center for
 Theoretical Physics, Universit\"at Bonn, Nussallee 14-16, D--53115 Bonn, Germany \label{addr2}
}

\date{}

\maketitle

\begin{abstract}

The Cottingham formula expresses the leading contribution of the electromagnetic interaction to the proton-neutron mass difference as an integral over the forward  Compton amplitude. Since quarks and gluons reggeize, the dispersive representation of this amplitude  requires a subtraction. We assume that the asymptotic behaviour is dominated by Reggeon exchange. This leads to a sum rule that expresses the subtraction function in terms of measurable quantities. The evaluation of this sum rule leads to $m_{\QED}^{p-n}=0.58\pm0.16\MeV$.  

\keywords{Dispersion relations \and Chiral Symmetries \and Electromagnetic mass differences \and Elastic and Compton scattering \and Protons and neutrons} 
\PACS{11.55.Fv \and 11.30.Rd \and 13.40.Dk  \and 13.60.Fz \and 14.20.Dh }
\end{abstract}

\tableofcontents

\section{Introduction}
\label{sec:Intro}
In the framework of the Standard Model, the fact that proton and neutron have nearly the same mass is explained as consequence of an approximate symmetry: isospin \cite{Heisenberg}. The symmetry is broken explicitly because the two lightest quarks neither have the same charge nor the same mass. The violation of the symmetry is very weak, because the e.m.~coupling $e$ as well as the difference between $m_u$ and $m_d$ are small. The weak interaction provides the neutron mass with an imaginary part and generates a shift of the real part as well, but these effects are tiny and will be neglected. The Standard Model then reduces to QED+QCD. In that framework, the expansion of the mass difference between proton and neutron in powers of $e$ starts with 
\be \label{eq:mpn} m^p-m^n=m_{\QCD}+m_{\QED}+O(e^4)\,,\ee 
where $m_{\QCD}$ is what remains if $e$ is turned off and is proportional to $m_u-m_d$, while $m_{\QED}$ stands for the term of order $e^2$. It is well-known that the splitting of the physical masses into an electromagnetic and a strong part is not unique. In our analysis, the ambiguity shows up through the scale of the logarithm occurring in the e.m.~renormalization of the quark masses and will be discussed in detail.

As shown by Cottingham \cite{Cottingham}, the leading contribution of the e.m.~interaction to the mass of a particle is given by an integral over the spin averaged forward Compton scattering amplitude, 
\be \label{eq:Compton} T^{\mu\nu}(p,q)=\text{\small $\frac{i}{2}$}\!\int\!\! d^4x\,e^{i q\cdot x}\langle p|Tj^\mu(x) j^\nu(0)|p\rangle\,.\ee
This amplitude is determined by QCD. If the mass of a particle
is expanded in powers of the e.m.~coupling constant $e$, the explicit expression for the term of order $e^2$ is formally given by\footnote{We fix the normalization of the one-particle states with\\ $\langle p',s'|p,s\rangle=2 p^0(2\pi)^3 \delta^3(\vec{p}'-\vec{p})\delta_{s' s}$ and $m$ is the mass of the particle. The spin averaged matrix element of the operator $O$ is abbreviated with  $\langle p|O|p\rangle \equiv\frac{1}{2}\sum_s \langle p,s|O|p,s\rangle$.}
\be\label{eq:mgamma} m_\gamma=\frac{ie^2}{2m(2\pi)^4}\int\!\!d^4q \,\frac{1}{q^2+i\epsilon} \,T^\mu_{\;\mu}(p,q)\,.\ee

There are two problems with this formula: (i) The short distance properties of QCD imply that the amplitude $T^\mu_{\;\mu}(p,q)$ does not fall off rapidly enough at large values of $q$ for the integral to converge.  (ii) $T^\mu_{\;\mu}(p,q)$ does not obey an unsubtracted dispersion relation -- causality alone determines the Compton amplitude through the structure functions of lepton-nucleon scattering only up to a subtraction function. The asymptotic behaviour in the deep inelastic region is now fully understood on the basis of asymptotic freedom, but the properties of the subtraction function are still under debate. 

Elitzur and Harari \cite{ElitzurHarari} pointed out that if the exchange of Reggeons correctly describes the asymptotic behaviour in the limit $\nu\rightarrow \infty$ at fixed $q^2$ -- an assumption we refer to as Reggeon dominance -- then the subtraction function obeys a sum rule which fully determines it through the cross section of lepton-nucleon scattering. Their paper appeared in 1970, at a time when the origin of the $\Delta I=1$ mass differences within the isospin multiplets was totally mysterious: evaluations of the Cottingham formula invariably led to the conclusion that the proton should be heavier than the neutron and hence unstable. 

In 1975 Gasser and Leutwyler  \cite{GL1975} then showed that the mystery disappears if the popular conviction, according to which the strong interaction conserves isospin, is dismissed. They showed that a coherent picture of isospin breaking can be reached within the Quark Model, provided the masses of the two lightest quarks are not only very small but also very different. At that time, the experimental results on deep inelastic scattering were consistent with the scaling laws of Bjorken \cite{Bjorken}. The implications of Reggeon dominance were worked out in this framework, using models to substitute the lack of experimental information in part of phase space, with the result $m_{\QED}=0.7\pm 0.3\MeV$ \cite{GL1975}. 

Lattice calculations of the proton-neutron mass difference are very demanding and became feasible only in the 21st century.  Early calculations were consistent with the result obtained from the Cottingham formula, but more recent evaluations indicate higher values for $m_{\QED}$ -- we will compare the available results with the outcome of our calculation in section \ref{sec:Lattice}. 

In 2012, Walker-Loud, Carlson and Miller \cite{WCM} performed a new evaluation of the Cottingham formula. They claimed that the analysis in \cite{GL1975} is inconsistent and replaced our sum rule by a model where the subtraction function $T_1(0,q^2)$ is parametrized with a simple algebraic formula. This paper triggered renewed interest and several authors investigated the matter \cite{ESTY,TWY,Walker-Loud2018,Tomalak}. We will discuss these works in section \ref{sec:Comparison}. A critical examination of some of the claims made in \cite{WCM} can be found in appendix E of \cite{GHLR} and in \cite{HLCD15,Hoferichter:2019jhr}.
 
The discovery of QCD and asymptotic freedom led to a fully transparent picture for the properties of the Compton amplitude in the region where both $\nu$ and $q^2$ are large and where the divergence of the Cottingham formula arises \cite{GrossWilczek,Politzer,Weinberg1973,Collins}. In Ref.~\cite{sumruleLetter}, we showed that the formal relation \eqref{eq:mgamma} can be rewritten in such a way that the divergences are under full theoretical control, exclusively concern the contribution from the subtraction function and are absorbed in the e.m.~renormalization of quark masses and QCD coupling constant. The aim of the present paper is to describe the analysis underlying these statements  in detail.   
 
The presentation is organized as follows. In a first part, sections \ref{sec:Lorentz invariance}--\ref{sec:Decomposition}, we discuss the mathematical underpinnings: decomposition of the Compton amplitude, dispersion relations, sum rule for the subtraction function, Wick rotation, mass formulae. The second part, sections \ref{sec:OPE}--\ref{sec:C}, deals with the operator product expansion, which governs the behaviour of the amplitudes at large momenta. The renormalization of the mass difference is discussed in sections \ref{sec:Renormalization} and \ref{sec:Subleading divergence}, whereas the data concerning the structure functions used in our work and the numerical determination of the subtraction function and of the mass difference are described in sections \ref{sec:Input}--\ref{sec:Numerics}.  
Sections \ref{sec:Lattice} and \ref{sec:Comparison} compare the outcome of our analysis with results obtained on the lattice and with other recent evaluations of the Cottingham formula. A summary and conclusions are provided in section \ref{sec:Summary}. The Appendices contain material concerning the operator product expansion as well as a detailed derivation of the sum rule for the subtraction function that plays a central role in the present work.  

\section{Lorentz invariance, kinematic zeros}
\label{sec:Lorentz invariance} 
Causality ensures that the time-ordered amplitude is unique up to contact terms and the ambiguity can be fixed in such a manner that $T^{\mu\nu}(p,q)$ is Lorentz cova\-riant.\footnote{Lorentz transformations and $P,T$ symmetry of the Compton amplitude are discussed in  detail in Appendix B of Ref. \cite{GHLR}.} Together with symmetry under space reflections, this property implies that the Compton amplitude can be decomposed as
\be \label{eq:Tmunu}T^{\mu\nu}(p,q)=A\, g^{\mu\nu}+B\, p^\mu p^\nu+C p^\mu q^\nu+C' p^\nu q^\mu+D\, q^\mu q^\nu\,,\ee
where $A,B,C,C',D$ only depend on the two variables $q^2$ and $\nu=p\cdot q/m$. Current conservation imposes the constraints
\be\label{eq:current conservation} A+ m\hspace{0.05em} \nu\, C+q^2D=0\,,\hspace{0.2em}
m\hspace{0.05em} \nu\,B+q^2C=0\,,\hspace{0.2em}C'=C\,.\ee
Since the physical spectrum of QCD does not contain massless particles, the amplitude $T^{\mu\nu}(p,q)$ cannot have a pole at $q^2=0$. Hence the second relation shows that $B$ vanishes for $q^2=0$ and can therefore be represented as $B=-q^2 T_2/m^2$. Setting $T_1=D$ and solving the constraints \eqref{eq:current conservation} for $A$ and $C$, this leads to the decomposition
\bea \label{eq:TK} \al\al T^{\mu\nu}(p,q)=T_1(\nu,q^2)K_1^{\mu\nu} +T_2(\nu,q^2)K_2^{\mu\nu}\,, \\
 \al\al K_1^{\mu\nu}= q^\mu q^\nu-g^{\mu\nu}q^2,\no
  \al\al K_2^{\mu\nu}= \frac{1}{m^2}\{(p^\mu q^\nu+p^\nu q^\mu)p\cdot q
 -g^{\mu\nu}(p\cdot q)^2-p^\mu p^\nu q^2\}.\nonumber
\eea
Crossing symmetry, $T^{\mu\nu}(p,q)=T^{\nu\mu}(p,-q)$, implies that $T_1$ and $T_2$ are even in $\nu$.

A popular alternative decomposition identifies the two independent amplitudes instead with $\hat{T}_1=-A$ and $\hat{T}_2=m^2B$. It is related to the one specified above by
\be\label{eq:That}\hat{T}_1=q^2T_1+\nu^2 T_2\,,\hspace{1em}\hat{T}_2=-q^2 T_2\,.\ee
The problem with this choice is that, in contrast to $T_1, T_2$, the amplitudes  $\hat{T}_1,\hat{T}_2$ contain kinematic zeros. This makes it difficult to determine their asymptotic behaviour. That is important because analytic functions are fully determined by their singularities only if the asymptotic behaviour is known. In dispersion theory, theoretical constraints are needed to determine the asymptotic properties of the amplitudes.   

To illustrate the problems encountered when working with amplitudes that are not free of kinematic zeros, consider the ``Born terms'', i.e.~the poles generated by the one-particle intermediate states. Their residues are determined by the elastic form factors of the nucleon. The Cauchy formula implies that an analytic function of the variable $z$ is determined uniquely by its singularities (poles, cuts) and by its behaviour for $z\rightarrow\infty$. The amplitudes $T_1$ and $T_2$ are analytic in $\nu$ at fixed $q^2$. The Born terms concern the contributions from the nucleon poles at $\nu=\pm \,q^2/2m$. They are fixed uniquely by the requirement that they disappear for $\nu\rightarrow\infty$ \cite{GHLR}:
\bea\label{eq:Born term}T_1^{\el}(q^2,\nu)\al=\al \frac{4m^2q^2\{G_E^2(q^2)-G_M^2(q^2)\}}{\{4m^2\nu^2-(q^2+i\epsilon)^2\}(4m^2-q^2)}\,, \\
T_2^{\el}(q^2,\nu)\al=\al -\frac{4m^2\{4m^2G_E^2(q^2)-q^2G_M^2(q^2)\}}{\{4m^2\nu^2-(q^2+i \epsilon)^2\}(4m^2-q^2)}\,.\nonumber\eea
For notation, in particular also for the definition of the Sachs form factors $G_E$ and $G_M$, we refer to \cite{GHLR}.

For the alternative decomposition \eqref{eq:That}, the elastic part of $\hat{T}_1$ does not disappear when $\nu\rightarrow\infty$.
In terms of Regge poles, the elastic part of $\hat{T}_1$ contains a fixed pole at $\alpha = 0$, with a residue that is determined by the nucleon form factors: $\hat{T}_1$ picks up asymptotic contributions that do not have anything to do with the phenomena that dominate the high energy behaviour of the Compton amplitude -- they merely reflect the fact that the amplitude $\hat{T}_1$ contains kinematic zeros. It is not advisable to work with such amplitudes -- for further discussion of the problems encountered in the presence of kinematic zeros, we refer to \cite{Bardeen,Tarrach,Hoferichter:2019jhr}.

As pointed out in the letter \cite{sumruleLetter}, the operator product expansion shows that, up to normalization, the leading spin 2 contributions in $T_1$ and $T_2$ are the same: in the combination
\be\label{eq:Tbar}\bar{T}(\nu,q^2)=T_1(\nu,q^2)+\mbox{$\frac{1}{2}$}T_2(\nu,q^2)\,,\ee 
these contributions drop out.  For this reason, the analysis of the asymptotic behaviour simplifies considerably if the pair $T_1$, $T_2$ is replaced by the pair $\bar{T}$, $T_2$, which is also free of kinematic zeros.

\section{Dispersion relations}
\label{sec:Dispersion relations}
The dispersion relations express the Compton amplitude in terms of the structure functions. These represent the Fourier transform of the current commutator:
\bea\label{eq:V}V^{\mu\nu}(p,q)\al=\al\frac{1}{4\pi}\!\int\!\!d^4 x e^{iq\cdot x}\langle p|[j^\mu(x),j^\nu(0)]|p\rangle \,,\\
V^{\mu\nu}(p,q)\al=\al V_1(\nu,q^2)K_1^{\mu\nu} +V_2(\nu,q^2)K_2^{\mu\nu}\,.\nonumber\eea
The structure functions are experimentally accessible only for $q^2\leq 0$ and it is customary to replace $q^2$ by $Q^2\equiv -q^2$. In the standard notation, where the structure functions are denoted by $F_1(x,Q^2)$, $F_2(x,Q^2)$ with $x=Q^2/2m\nu$, $V_1$ and $V_2$ are given by:
\bea\label{eq:V12} V_1\al=\al\frac{F_L}{2x Q^2}\,,\hspace{1em}V_2=\frac{F_2}{2x \nu^2}\,,\\
F_L\al=\al F_2-2x F_1\,.\nonumber\eea
For $\bar{T}$, the structure function $\bar{V}=V_1+\frac{1}{2}V_2$ is relevant:
\bea\label{eq:Vbar} \bar{V}\al=\al  \frac{\bar{F}}{2x Q^2}\,,\\
 \bar{F}\al=\al F_L+\frac{2m^2 x^2}{Q^2}F_2\,.\nonumber\eea

We assume that the Compton amplitude exhibits Regge behaviour for $\nu\rightarrow\infty$: $\bar{T}\propto\nu^\alpha$, $T_2\propto\nu^{\alpha-2}$. Accordingly, the dispersion relation for $\bar{T}$ requires a subtraction while $T_2$ obeys an unsubtracted dispersion relation:
\bea\label{eq:DRTbar}\bar{T}(\nu,q^2)\al=\al\bar{S}(q^2)+\bar{T}^{\el}(\nu,q^2)\\
\al\al\hspace{-1em} +2(\nu^2-\nu_0^2)\!\int_{\nuth}^\infty\hspace{-0.9em} \nu'd\nu'\frac{\bar{V}(\nu',q^2)}{(\nu'^2-\nu_0^2)(\nu'^2-\nu^2-i\epsilon)}\,,\no
\label{eq:DRT2}T_2(\nu,q^2)\al=\al T_2^{\el}(\nu,q^2)+2\!\int_{\nuth}^\infty\hspace{-0.9em}d\nu' \nu' \frac{ V_2(\nu',q^2)}{\nu'^2-\nu^2-i\epsilon}\,. 
\eea
$\bar{S}(q^2)$ represents the subtraction function, $\nu_0^2$ is the subtraction point in the variable $\nu^2$ and the lower limit corresponds to the threshold for inelastic reactions, $\nuth=(2m M_\pi+M_\pi^2-q^2)/(2m)$. The elastic part of $\bar{T}$ is given by $\bar{T}^{\el}=T_1^{\el}+\frac{1}{2}T_2^{\el}$.  

As such, the choice of the subtraction point is arbitrary (provided that $\nu_0^2<\nuth^2$), but as pointed out in \cite{sumruleLetter}, it is convenient to set $\nu_0^2=-\frac{1}{4}Q^2$ rather than to subtract at $\nu_0=0$. As will be seen below, this choice simplifies the asymptotic behaviour of the subtraction function for $Q^2\rightarrow\infty$. Replacing the variable of integration $\nu'$ by $x=Q^2/(2m \nu')$, the dispersive representation then takes the form
\bea\label{eq:DRx}\bar{T}(\nu,-Q^2)\al=\al\bar{S}(-Q^2)+\bar{T}^{\el}(\nu,-Q^2) \\
\al\al\hspace{-5.7em}+\, ( Q^2+ 4\nu^2)\!\int_0^{\xth}\hspace{-1.3em}dx\, \frac{m^2 \bar{F}(x,Q^2)}{(Q^2+m^2 x^2)
(Q^4-4m^2 x^2 \nu^2-i\epsilon)}\,,\no
T_2(\nu,-Q^2)\al=\al T_2^{\el}(\nu,-Q^2)+\!\int_{0}^{\xth}\hspace{-1.3em}dx\frac{4m^2F_2(x,Q^2)}{Q^4-4m^2 x^2\nu^2-i\epsilon} \,.\nonumber\eea
with $\xth=Q^2/(Q^2+2m M_\pi +M_\pi^2)$. 

\section{Reggeon dominance}
\label{sec:Reggeon dominance}
While $T_2$ is fully determined by the form factors and the structure functions because it obeys an unsubtracted dispersion relation, the representation for $\bar{T}$ involves a subtraction function, which causality alone leaves undetermined. This illustrates a venerable theorem which concerns the implications of causality for the structure functions \cite{JostLehmann,Dyson,BogoliubovVladimirov}. The theorem states that the values of  $\bar{V}(\nu,q^2)$, $V_2(\nu,q^2)$ in the space-like region $q^2\leq 0$ determine these functions in the time-like region, up to a polynomial in the variable $\nu$. The implications for the dispersive analysis of the Compton amplitude are discussed in \cite{LO}.

In Regge language, integer powers of $\nu$ are called fixed poles: the continuation from the space-like to the time-like region is unique up to fixed poles. Regge asymptotics excludes such contributions in $V_2$, but the continuation of $\bar{V}$ into the time-like region is unique only up to a term that depends on $\nu$ exclusively through the step function: 
\be\bar{V}^{\fp}(\nu,q^2)=\epsilon(\nu)\sigma(q^2)\,,\ee
where $\sigma(s)$ vanishes for $s<0$. In $\bar{T}$, the ambiguity shows up in the form
\be \bar{T}^{\fp}(q^2)=\int_0^\infty \!\! ds\,\frac{\sigma(s)}{s-q^2-i\epsilon}\,,\ee
which is independent of $\nu$ and thus only affects the subtraction function. Since the ambiguity amounts to a superposition of free propagators, it is evident that the Fourier transform of $\bar{V}^{\fp}$ vanishes outside the light-cone. The nontrivial part of the theorem is that the values of the structure functions in the space-like region, where they can be measured, fully determine the amplitude in the time-like region up to contributions of this particular type.  
 
In QED, the electrons reggeize \cite{Gell-Mann1964}, but the photon remains elementary \cite{Mandelstam1965}. QCD, however, does satisfy the criteria for Reggeization formulated in \cite{Mandelstam1965}:  the gluons  as well as the quarks have this property \cite{Grisaru1,Grisaru}. In the meantime, the graphs that need to be summed up to study the high energy properties of the scattering amplitudes within QCD perturbation theory have been identified and Reggeon field theory has been developed for the analysis of exchanges of more than one Reggeon, in particular also of the cuts generated by these  \cite{Lipatov,Kuraev,Balitsky,Gribov2003}. 

It is generally assumed that the asymptotic behaviour of the Compton amplitude is indeed governed by Reggeon exchange. The exchange of Reggeons generates contributions which at high energies are of the form
\be\label{eq:TR} \bar{T}^{\indR}(\nu,q^2)=-\sum_{\alpha>0}\frac{\pi\beta_\alpha(q^2)}{\sin\pi\alpha}\{(-s)^{\alpha}+(-u)^\alpha\}\,,\ee
where $s =m^2+ 2m \nu+q^2$ and $u=m^2-2m \nu+q^2$ represent the square of the centre of mass energy in the $s$- and $u$-channels, respectively. In general, the power $\alpha$ depends on $t$: $\alpha(t)$ moves on a Regge trajectory. In our context, however, only the forward scattering amplitude is of interest, so that only the intercept $\alpha=\alpha(0)$ is relevant. For the Compton amplitude of the proton or the neutron, the Pomeron yields the dominant contribution; it involves a superposition of terms of the above form with intercepts in the vicinity of 
$\alpha=1$. The Reggeon with the quantum numbers $I^C=1^+$ and an intercept in the vicinity of $\alpha=\frac{1}{2}$ represents the most important non-leading contribution. We refer to this Reggeon as the $a_2$. In the difference between the amplitudes relevant for proton and neutron, the Pomeron drops out -- the $a_2$ represents the leading term. 

We assume that the Reggeons dominate the asymptotic behaviour \cite{GL1975}:
\be\label{eq:RD}\lim\rule[-0.7em]{0em}{1em}_{\hspace{-1.7em}\nu\rightarrow\infty} (\bar{T}-\bar{T}^{\indR})= 0\,.\ee
This amounts to the assumption that the amplitude $\bar{T}$ does not contain a fixed pole at $\alpha=0$. We do not know of a physical phenomenon that could produce a fixed pole at $\alpha=0$  in $\bar{T}$. Neither causality nor the short-distance singularities nor the Reggeons generate terms of this sort.

The constraints imposed on the subtraction function by causality and unitarity have been analyzed within the alternative dispersive framework set up in \cite{Caprini}. Model-independent bounds for the subtraction function $S_1(q^2)$ are derived and it is shown that the results obtained  in \cite{GHLR} from Reggeon dominance at low values of $Q^2$ are consistent with these. An extension of this work to the higher values of $Q^2$ investigated in the present paper would be most welcome, as it would allow to subject Reggeon dominance to a further test. Model-independent bounds on the subtraction function $\bar{S}(q^2)$ would be particularly interesting, because the operator product expansion of this quantity is free of the short distance singularities associated with operators of spin 2 -- asymptotic freedom fully determines the asymptotic behaviour of $\bar{S}$. 
\section{Sum rule for the subtraction function}
\label{sec:Sum rule}
As pointed out in \cite{GL1975}, Reggeon dominance determines the subtraction function in terms of the cross sections of inelastic scattering. In \cite{GHLR}, a sum rule for the subtraction function $S_1(-Q^2)\equiv T_1(0,-Q^2)$ was derived that represents the inelastic part of this quantity in terms of integrals over the cross sections. We now derive an analogous sum rule that expresses the subtraction function $\bar{S}(-Q^2)$ as an integral over the structure function $\bar{F}(x,Q^2)$ -- an immediate consequence of the dispersion relation \eqref{eq:DRx} and Reggeon dominance \eqref{eq:RD}. The derivation is not trivial, however, because the Reggeons generate singularities at $x=0$. The leading singularity in $\bar{F}$ is of the form:
\bea\label{eq:FR}\al\al\bar{F}^{\indR}(x,Q^2)= \sum_{\alpha>0} b_\alpha(Q^2)x^{1-\alpha}\,,\\
\al\al b_\alpha(Q^2)= 2\hspace{0.05em}Q^{2(\alpha+1)}\beta_\alpha(-Q^2)\,.\nonumber\eea 
For this reason, taking the limit $\nu\rightarrow\infty$ in the dispersion relation \eqref{eq:DRx} requires some care: the limit cannot simply be exchanged with the dispersion integral
\bea\label{eq:Tbardisp}
\al\al \bar{T}^{\disp}(\nu,-Q^2)= \\
\al\al(Q^2+ 4\nu^2)\!\int_0^{\xth}\hspace{-1.3em}dx\, \frac{m^2 \bar{F}(x,Q^2)}{(Q^2+m^2 x^2)
(Q^4-4m^2 x^2 \nu^2-i\epsilon)}\,,\nonumber\eea
because, in the vicinity of $x=0$, the term $4m^2 x^2\nu^2$ fails to dominate over $Q^4$.  Since the elastic part of the amplitude tends to zero for $\nu\rightarrow\infty$, the Reggeon dominance hypothesis \eqref{eq:RD} amounts to the requirement that the subtraction function cancels the limiting value of  the difference between the dispersion integral and the asymptotic representation: 
\be\label{eq:Sbarlim} \bar{S}=-\lim\rule[-0.7em]{0em}{1em}_{\hspace{-1.7em}\nu\rightarrow\infty} (\bar{T}^{\disp}-\bar{T}^{\indR})\,.\ee
The limit is worked out in Appendix \ref{sec:Derivation}. The result takes the form of a sum rule that expresses the subtraction function $\bar{S}$ in terms of the structure function $\bar{F}$ 
\bea\label{eq:Sum rule}Q^2\bar{S}(-Q^2)\al=\al\int_0^{\xth}\hspace{-1.3em}dx\,\frac{\bar{F}(x,Q^2)-\bar{F}^{\indR}(x,Q^2)}{x^2}- \sum_{\alpha>0}\frac{b_\alpha(Q^2)}
{\alpha\, \xth^\alpha}\no \al-\al m^2\hspace{-0.4em} \int_0^{\xth}\hspace{-1.3em}dx\, \frac{\bar{F}(x,Q^2)}{Q^2+m^2 x^2}\,.\eea
A Finite Energy Sum Rule variant of this relation was proposed by Elitzur and Harari \cite{ElitzurHarari}. The above formulation shows that the sum rule is perfectly consistent with the scaling violations required by QCD -- contrary to statements made in \cite{Collins}.  
 
\section{Wick rotation}
\label{sec:Wick rotation}
We now return to the mass formula \eqref{eq:mgamma}.  In the decomposition introduced above, the trace of the Compton amplitude is given by
\be\label{eq:Tmumu}T^\mu_{\;\mu}(p,q)=-\{3 q^2\bar{T}(\nu,q^2)+2(\nu^2-\mbox{$\frac{1}{4}$}q^2)T_2(\nu,q^2)\}\,.\ee
The integration in (\ref{eq:mgamma}) runs over all $q^2$, space-like as well as time-like. For the integral to converge, it needs to be regularized, for instance by replacing the photon propagator $1/q^2$ with $\Lambda^2/(\Lambda^2-q^2)/q^2$. 

In the rest frame of the particle, $\nu$ coincides with the component $q^0$ of the photon momentum.
 Cottingham \cite{Cottingham} observed that the time-ordered amplitude is analytic in $q^0$ and that the path of integration over this variable may be  rotated into the imaginary axis, at fixed three momentum $\boldsymbol{q}$ -- without crossing any singularities of the integrand in \eqref{eq:mgamma} (Wick rotation). 
 
 Setting $q^0=i Q_4$ and identifying $Q_1,Q_2,Q_3$ with the space components of the physical momentum, $m_\gamma$ takes the form  of a euclidean integral extending over the four-vector $Q_\mu$:
\be \label{eq:mgammaLambda1} m_\gamma=\frac{e^2}{2m(2\pi)^4}\hspace{-0.3em}\int \hspace{-0.2em}\frac{d^4Q}{Q^2}\frac{\Lambda^2}{\Lambda^2+Q^2}\{3Q^2\bar{T}+2(Q_4^2-\mbox{$\frac{1}{4}$}Q^2)T_2\}\,. \ee
The result  for the renormalized mass difference is independent of the form used for the regularization. It is customary to use a cutoff in momentum space: restrict the integration to the euclidean sphere $Q^2\leq \Lambda^2$ and write the regularized Cottingham formula as
\be \label{eq:mgammaLambda} m_\gamma^\Lambda=\frac{e^2}{2m(2\pi)^4}\hspace{-0.3em}\int\hspace{-2em}\rule[-1.5em]{0em}{0em}_{Q^2\leq \Lambda^2}\hspace{-0.8em}\frac{d^4Q}{Q^2}\{3Q^2\bar{T}+2(Q_4^2-\mbox{$\frac{1}{4}$}Q^2)T_2\}\,.\ee
In this formula, the amplitudes $\bar{T}$ and $T_2$ are to be evaluated  at $\nu=iQ_4$, $q^2=-Q^2$. 
 
\section{Decomposition of the mass shift}
\label{sec:Decomposition}

In the framework of QCD+QED, the mass of a particle is determined by the bare parameters that occur in the Lagrangian and the cutoff used to regularize the theory. If the electromagnetic interaction is turned off, only the QCD coupling constant, the quark masses and the cutoff are relevant. To order $e^2$, the e.m. interaction changes the mass not only by the above integral, but in addition by the contribution $\Delta m^\Lambda$, which arises from the change in the bare parameters needed for the mass of the particle to stay finite when the cutoff $\Lambda$ is removed -- the bare quantities depend on $\Lambda$ as well as on $e$. 
The e.m. part of the mass is obtained by adding this contribution to the integral in equation \eqref{eq:mgammaLambda}:
\be\label{eq:lim} m_{\QED}= \mbox{lim}\hspace{-1.8em}\rule[-0.8em]{0em}{0em}_{\Lambda\rightarrow\infty}\{m_\gamma^\Lambda+\Delta m^\Lambda\}\,.\ee

Inserting the dispersion relations \eqref{eq:DRx} in formula \eqref{eq:mgammaLambda}, we obtain a representation of the e.m.~part of the mass as a sum of four terms:
 \be\label{eq:decomposition}  m_{\QED}= m_{\el}+m_{\bar{F}}+m_{F_2}+m_{\bar{S}}\,.\ee
In each one of these, the integrals over the direction of the vector $Q_\mu$ can be done explicitly.  

In the first term, which collects the contributions from $\bar{T}^{\el}$ and $T_2^{\el}$, this leads to a set of integrals over the form factors of the nucleon, which are known very accurately -- an explicit expression for $m_{\el}$ is given, for instance, in \cite{GHLR}. 

The second and third term arise from the dispersion integrals over the structure functions $\bar{F}$ and $F_2$, respectively. With the above choice of the subtraction point, the integrands are proportional to the factor $Q^2-4Q_4^2$, in both cases. Taken by itself, the angular integral of this factor over the directions of the vector $Q_\mu$ vanishes. Moreover, when $Q^2$ becomes large, the remainder of the integrand becomes independent of $Q_4$. Hence the angular integration suppresses the contributions arising from large values of $Q^2$ -- these integrals approach finite limits when the cutoff is removed: 
\bea\label{eq:mFbar}m_{\bar{F}}\al=\al 4N\!\!\int_0^\infty\hspace{-1em} dQ^2 \!\int_0^{\xth}\hspace{-1.3em}dx \; \frac{yf(y)\bar{F}(x,Q^2)}{x^2(1+4y)}\,,\\ \label{eq:mF2}
m_{F_2}\al=\al -\mbox{$\frac{2}{3}$}N\!\!\int_0^\infty\hspace{-1em} dQ^2\!\int_0^{\xth}\hspace{-1.3em}dx\;\frac{f(y)F_2(x,Q^2)}{x^2}\,.\eea
The normalization constant is given by 
\be\label{eq:N}N=\frac{3\alphaem}{8\pi m}\,,\ee
the variable $y$ stands for $y=Q^2/(4m^2 x^2)$  and the explicit expression for the function $f(y)$ reads
\be\label{eq:f(y)} f(y)= \frac{1+4y}{2}\sqrt{1+\frac{1}{y}}-\frac{3+4y}{2}\,.\ee
The suppression of the angular integrals relevant for $m_{\bar{F}}$ and $m_{F_2}$ manifests itself in the fact that the function $f(y)$ rapidly falls off when $y$ becomes large:
\be f(y)=\frac{1}{16y^2}+O(y^{-3})\,.\ee

The angular integral can be done explicitly in the fourth term as well, but there, it does not suppress the contributions from large values of $Q^2$, so that the cutoff must be retained: 
\bea\label{eq:mS}m_{\bar{S}}\al=\al \mbox{lim}\hspace{-1.8em}\rule[-0.8em]{0em}{0em}_{\Lambda\rightarrow\infty}\left\{N\!\!\int_0^{\Lambda^2}\hspace{-1em} dQ^2 Q^2 \bar{S} (-Q^2)+\Delta m^\Lambda\right\}\,.\eea

Together with the sum rule \eqref{eq:Sum rule}, the above formulae fully specify the e.m.~part of the mass difference between proton and neutron, in terms of measurable quantities. The next four sections concern asymptotic properties of the Compton amplitude that are not of direct relevance for the Cottingham formula -- the contributions generated by short distance singularities of spin 2, for instance. If the reader is more interested in the numerical outcome of our analysis for the mass difference, he or she may go directly to section \ref{sec:Moments}.

\section{Operator product expansion}
\label{sec:OPE} 
The behaviour of the amplitudes $\bar{T}$ and $T_2$ at large momenta is determined by the short-distance properties of the matrix element $\langle p|Tj^\mu(x) j^\nu(0)|p\rangle$, which can be analyzed by means of the operator product expansion (OPE) \cite{Wilson}. The asymptotic freedom of QCD implies that perturbation theory can be used to work out the leading terms of this expansion \cite{GrossWilczek,Politzer,Weinberg1973,Collins}. For the time-ordered product of two currents, the behaviour at short distances $z=x-y$ is of the form
\be\label{eq:Tmunuz} Tj_\mu(x) j_\nu(y)\;\rightarrow\hspace{-1.55em}\rule[-0.6em]{0em}{0em}_{z\rightarrow 0}\hspace{0.2em}\sum_n 
\tilde{C}_{\mu\nu}^n(z)O_n(X)\,,\ee
where $O_n$ enumerates the renormalized gauge invariant local operators of QCD and $X=\frac{1}{2}(x+y)$. The expansion starts with the operators of lowest dimension. The Wilson coefficients $\tilde{C}_{\mu\nu}^n(z)$ vanish unless $O_n$ has the same flavour quantum numbers as the product of two e.m.~currents. The symmetry of QCD under P, T and C also prevents some operators from contributing to the expansion. The coefficients depend in a nontrivial manner  on $z$ only through $z^2$: they are polynomials in the components  of the vector $z$, with coefficients that depend on $z^2$.

In momentum space, the OPE governs the behaviour at large momenta. We denote the Fourier transform with respect to $z$ by 
$\tilde{T}_{\mu\nu}$:
\be\label{eq:TtildeX} \tilde{T}_{\mu\nu}(q,X)=\text{\small $\frac{i}{2}$}\!\int\!\! d^4z\,e^{i q\cdot z} Tj_\mu(x) j_\nu(y) \,.\ee
The limit $z=\lambda \bar z$, $\lambda\to 0$ in coordinate space corresponds to the limit $q=\lambda \bar q$, $\lambda\to \infty$ in momentum space.
We refer to this limit as $q\rightarrow\infty$. In this notation, we have
\be\label{eq:ThatC}\tilde{T}_{\mu\nu}(q,X) \hspace{0.5em}\rightarrow\hspace{-1.7em}\rule[-0.6em]{0em}{0em}_{q\rightarrow \infty}\hspace{0.2em}
 \sum_n C_{\mu\nu}^n(q) O_n(X) \,.\ee
The coefficients $C_{\mu\nu}^n(q)$ represent the Fourier transforms of  those in equation \eqref{eq:Tmunuz} and are polynomials in the components of $q$, with coefficients that depend on $q^2$. This immediately implies that the coefficients occurring in the expansion of the invariant amplitudes $T_1(\nu,q^2)$, $T_2(\nu,q^2)$ are polynomials in $\nu$. The expansion thus also holds for imaginary values of $\nu$.  

A contribution from the unit operator only occurs in the disconnected part and does not show up in the scattering amplitude. In QCD, the relevant operators of lowest dimension are $\fbar f$ with $f=u, d, \ldots$ They have spin zero and are of engineering dimension 3. Chiral symmetry, however, suppresses the contributions from these operators: their Wilson coefficients are proportional to the masses $m_u, m_d,\ldots$  It is convenient to include the mass factor and to work with the operator $O^{f_0}=m_f\fbar f$, which is of dimension 4. Lorentz invariance implies that the spin averaged matrix elements of the operators of lowest dimension can be expressed in terms of the following linearly independent operators, which either have spin 0 or spin 2:
\bea\label{eq:operators}
O^{f_0}\al=\al m_f \fbar f\,,\\
O^{f_2}_{\alpha\beta}\al=\al \mbox{$\frac{1}{2}$}i\fbar\{\gamma_\alpha D\hspace{-0.8em}\mbox{\scriptsize\raisebox{1em}{$\leftrightarrow$}}_\beta +\gamma_\beta D\hspace{-0.8em}\mbox{\scriptsize\raisebox{1em}{$\leftrightarrow$}}_\alpha-\mbox{$\frac{1}{2}$}g_{\alpha\beta}\gamma^\lambda  D\hspace{-0.8em}\mbox{\scriptsize\raisebox{1em}{$\leftrightarrow$}}_\lambda\}f\,,\no
O^{G_0}\al=\al \text{tr}\{G_{\mu\nu}G^{\mu\nu}\}\,,\no
O^{G_2}_{\alpha\beta}\al=\al \text{tr}\{G_{\alpha\lambda}G_{\beta}^{\hspace{0.4em}\lambda}\}-\mbox{$\frac{1}{4}$}g_{\alpha\beta}\text{tr}
\{G_{\lambda\mu}G^{\lambda\mu}\}\,.
\nonumber\eea
Accordingly, the leading terms in the operator product expansion of $\tilde{T}_{\mu\nu}(q,X)$ are given by
\bea\label{eq:ThatC1}\al\al\tilde{T}_{\mu\nu}(q,X)\hspace{0.5em}\rightarrow\hspace{-1.7em}\rule[-0.6em]{0em}{0em}_{q\rightarrow \infty}\hspace{0.2em}
 \sumf \left\{C_{\mu\nu}^{f_0}  O^{f_0}   +C_{\mu\nu}^{f_2\,\alpha\beta} O^{f_2}_{\alpha\beta} \right\}\\
\al\al\hspace{7.5em}+\,C_{\mu\nu}^{G_0}O^{G_0}+C_{\mu\nu}^{G_2\,\alpha\beta}O^{G_2}_{\alpha\beta}\,.\nonumber\eea
While the dependence on $q$ resides in the Wilson coefficients, the operators only depend on $X$.

\section{Leading Wilson coefficients}
\label{sec:Leading Wilson coefficients}
Lorentz invariance and current conservation imply that the Wilson coefficient of the scalar operator $O^{f_0}$ is proportional to the kinematic tensor $K_{1\,\mu\nu}$ specified in equation \eqref{eq:TK}: the contribution from this operator is of the form
\be\label{eq:Cf0} C_{\mu\nu}^{f_0}\,O^{f_0}=K_{1\,\mu\nu}\,c_1^{f}O^{f_0}\,,\ee
where the coefficient $c_1^f$ depends on $q^2$. The contribution from $O^{G_0}$ is of the same structure.

For operators with spin 2, the situation is not that simple. As shown in Appendix \ref{sec:Wilson2}, Lorentz invariance and current conservation determine the form of their Wilson coefficients only up to two functions of $q^2$, which we denote by $c_2(q^2)$ and $c_3(q^2)$. According to equation \eqref{eq:Cc23}, the contribution from $O_{\alpha\beta}^{f_2}$ is of the form:
\bea\label{eq:Cf2}\al\al C_{\mu\nu}^{f_2\,\alpha\beta}O_{\alpha\beta}^{f_2}=c_3^f\,(q_\mu q_\nu-g_{\mu\nu}q^2) O_{\alpha\beta}^{f_2}q^\alpha q^\beta \\
\al\al\hspace{1em} +\,c_2^f\, (g_{\mu\alpha}O_{\nu\beta}^{f_2}+g_{\nu\alpha} O_{\mu\beta}^{f_2}-g_{\mu\nu}O_{\alpha\beta}^{f_2}) (q^\alpha q^\beta-\mbox{$\frac{1}{2}$}g^{\alpha\beta})\,,\nonumber \eea
and analogously for the contribution from the lowest dimensional gluonic operator of spin 2.  This shows that kinematics determines the Wilson coefficients of the lowest dimensional operators in terms of six functions, $c_1^{f},\ldots \,,c^{G}_3$, that depend on $q^2$. 

The amplitude we are interested in represents the spin average of the one-particle matrix element
\be \label{eq:TThat}T_{\mu\nu}(p,q)=\langle p |\tilde{T}_{\mu\nu}(q,X)|p\rangle\,.\ee
(Since the initial and final momenta are the same, the matrix element is independent of $X$). Inserting the expansion \eqref{eq:ThatC1}, we obtain:
\bea\label{eq:TmunuC} T_{\mu\nu}(p,q)\al\al\hspace{0.5em}\rightarrow\hspace{-1.7em}\rule[-0.6em]{0em}{0em}_{q\rightarrow \infty}\hspace{0.2em}
 \sumf \left\{C_{\mu\nu}^{f_0}\langle p| O^{f_0}|p\rangle +C_{\mu\nu}^{f_2\,\alpha\beta}\langle p|O^{f_2}_{\alpha\beta}|p\rangle\right\}\no
\al\al\hspace{1.5em}+\,C_{\mu\nu}^{G_0}\langle p| O^{G_0}|p\rangle+C_{\mu\nu}^{G_2\,\alpha\beta}\langle p| O^{G_2}_{\alpha\beta}|p\rangle\,.
\eea

For scalar operators, the matrix element is a constant, while for spin 2, it depends on the momentum of the particle: 
\bea\label{eq:O2} \langle p|O^{f_0}|p\rangle \al=\al \langle O^{f_0}\rangle\,,\\
\langle p| O_{\alpha\beta}^{f_2}|p\rangle\al=\al(\hat{p}_\alpha \hat{p}_\beta -\mbox{$\frac{1}{4}$}g_{\alpha\beta})\langle O^{f_2}\rangle\,,\hspace{1em}\hat{p}_\alpha=\frac{p_\alpha}{m}\,.\nonumber\eea 

According to equation \eqref{eq:Cf0}, the contributions from the spin zero operators are proportional to the kinematic tensor $K_{1\,\mu\nu}$. For the spin 2 terms, a short calculation is needed to verify that it can be expressed as a linear combination of $K_{1\,\mu\nu}$ and 
$K_{2\,\mu\nu}$:
\bea C_{\mu\nu}^{f_2\,\alpha\beta}\langle O^{f_2}_{\alpha\beta}\rangle
\al = \al(-\mbox{$\frac{1}{2}$}K_{1\,\mu\nu}+K_{2\,\mu\nu})\,c_2^{f}\langle O^{f_2}\rangle\no
 \al + \al K_{1\,\mu\nu}(\nu^2-\mbox{$\frac{1}{4}$}q^2)\,c_3^{f}\langle O^{f_2} \rangle\,.\eea
The terms arising from the lowest dimensional gluonic operator of spin 2 are of the same form.

The corresponding asymptotic representations for $T_1(\nu,q^2)$  and $T_2(\nu,q^2)$ are given by the coefficients of $K_{1\,\mu\nu}$ and $K_{2\,\mu\nu}$, respectively:
\bea\label{eq:T1as}T_1(\nu,q^2)\al\hspace{0.2em}\rightarrow\hspace{-1.7em}\rule[-0.6em]{0em}{0em}_{q\rightarrow \infty}\hspace{0em} \al \sumf c_1^{f}(q^2)\langle O^{f_0}\rangle +c_1^{G}(q^2)\langle O^{G_0}\rangle\\
 \al\al\hspace{-2.5em} -\mbox{$\frac{1}{2}$} \sumf c_2^{f}(q^2)\langle O^{f_2}\rangle- \mbox{$\frac{1}{2}$}c_2^{G}(q^2)\langle O^{G_2}\rangle\no
\al\al \hspace{-2.5em}+(\nu^2-\mbox{$\frac{1}{4}$}q^2) \{\sumf c_3^{f}(q^2)\langle O^{f_2}\rangle +c_3^{G}(q^2)\langle O^{G_2}\rangle \}\,,\no 
\label{eq:T2as}T_2(\nu,q^2)\al\hspace{0.5em}\rightarrow\hspace{-1.7em}\rule[-0.6em]{0em}{0em}_{q\rightarrow \infty}\hspace{0.2em} \al  
 \sumf c_2^f (q^2)\langle O^{f_2}\rangle+c_2^{G}(q^2)\langle O^{G_2}\rangle\,.\eea
While the leading term in the asymptotic behaviour of $T_2(\nu,q^2)$ only depends on $q^2$, $T_1(\nu,q^2)$ contains a term proportional to $\nu^2$. 

The advantage of working with the amplitude $\bar{T}$ now becomes visible: $T_1$ and $T_2$ contain a common spin 2 contribution. In the combination $\bar{T}=T_1+\frac{1}{2}T_2$, this term drops out -- only the one proportional to the factor $\nu^2-\frac{1}{4}q^2$ remains:
\bea\label{eq:Tbaras}\bar{T}(\nu,q^2)\al\hspace{0.5em}\rightarrow\hspace{-1.7em}\rule[-0.6em]{0em}{0em}_{q\rightarrow \infty}\hspace{0.2em} \al   \sumf c_1^{f}(q^2)\langle O^{f_0}\rangle+ c_1^{G}(q^2)\langle O^{G_0}\rangle\\
\al\al\hspace{-4em}+(\nu^2-\mbox{$\frac{1}{4}$}q^2)\{ 
\sumf c_3^{f}(q^2)\langle O^{f_2}\rangle +c_3^{G}(q^2)\langle O^{G_2}\rangle\}\,.\nonumber\eea
As noted above, the angular integration suppresses contributions that are proportional to this factor -- this is the reason why in the decomposition \eqref{eq:decomposition} only $m_{\bar{S}}$ contains a divergence.  

\section{Difference  between proton and neutron}
\label{sec:Difference between  proton and nejutron}
For the mass difference between proton and neutron, only the difference between the Compton amplitudes of the two particles is needed. As far as the asymptotic behaviour is concerned, we thus only need the difference between the spin averaged matrix elements of proton and neutron. In the isospin limit, the neutron matrix elements of the gluonic operators $O^{G_0}$ and $O^{G_2}_{\alpha\beta}$ coincide with those of the proton. In reality, since $m_u$ differs from $m_d$, the proton and neutron matrix elements of the gluonic operators are slightly different, but in the mass difference between proton and neutron, this generates an effect of second order in isospin breaking and will be neglected.  
This simplifies matters considerably. Only the matrix elements of non-singlet operators are relevant -- operator mixing does not affect these.

Throughout the remainder of this paper, we focus on the difference between the Compton amplitudes of proton and neutron, without explicitly
indicating this in the notation: in the following, $\bar{T}$ and $T_2$ stand for $\bar{T}^{p-n}$ and $T_2^{p-n}$, respectively. 

\section{Perturbation theory}
\label{sec:Perturbation theory}
To leading order of the QCD perturbation series, the Wilson coefficients are the same as for free quarks. The explicit expressions are readily obtained by simply replacing the nucleon in the above relations with a free quark of charge  $Q_f=\frac{2}{3}$ or $-\frac{1}{3}$. If the strong interaction is turned off and the e.m.~interaction is accounted for only to leading order, the Compton scattering on a quark is elastic. The Sachs form factors are given by $G_E =  G_M = Q_f$, so that the formulae \eqref{eq:Born term} reduce to
\be T_1^f=0\,,\hspace{1em} T_2^f=\frac{4m^2_fQ_f^2}{(q^2+i\epsilon)^2-4m^2\nu^2}\,.\ee
 In the limit  $q\rightarrow \infty$ relevant for the OPE,  the second term in the denominator becomes negligible compared to the first:  for space-like momenta, $T_2^f$ tends to $4m_f^2Q_f^2/Q^4$. 
 The spin averaged quark matrix elements of the operators $O^{f_0}$ and $Q_{\alpha\beta}^{f_2}$ are readily worked out; they yield $\langle O^{f_0}\rangle=2m_f^2$ and $\langle O^{f_2}\rangle=4m_f^2$. The leading terms in the expansion of the coefficients  $c_1^f, c_2^f,c_3^f$ in powers of $g^2$ can then be read off from the asymptotic relations \eqref{eq:T1as} and \eqref{eq:T2as} \cite{Collins,HillandPaz}: 
\bea\label{eq:cf} c_1^f(-Q^2)\al=\al\frac{Q_f^2}{Q^4}+O(g^2)\,,\\
c_2^f(-Q^2)\al=\al \frac{Q_f^2}{Q^4}+O(g^2)\,,\no
c_3^f(-Q^2)\al=\al O(g^2)\,.\nonumber\eea
In this calculation, the spin of the operators occurring in the OPE does not play an important role. Appendix \ref{sec:OPE for free quarks} contains an alternative derivation of these relations, which is based on the short distance expansion of the quark propagator and explicitly exhibits the spin structure.
 
The higher order terms in the expansion of the Wilson coefficients have been studied in detail, also for the gluonic operators \cite{GrossWilczek,Politzer,Weinberg1973,Collins,HillandPaz} -- for a thorough review, we refer to \cite{Buras}. The qualitative features of the asymptotic structure are intimately related to the fact that the dimension of the spin 2 operators is anomalous.  The correction of order $g^2$ in the perturbative expansion of the quantity $Q^4c_2^f(-Q^2)$ falls logarithmically if $Q^2$ becomes large. With the renormalization group, the leading logarithms can be summed up to all orders. The contributions from the singlet operators undergo mixing, but as noted above, for the difference between proton and neutron, only the nonsinglet operators are relevant. The matrix element of the term involving $c_2^f$ falls off with
\be\label{eq:c2f} \sum_f c_2^f(-Q^2)\langle O^{f_2}\rangle^{p-n}\hspace{0.3em}\rightarrow\hspace{-1.9em}\rule[-0.6em]{0em}{0em}_{Q\rightarrow \infty}\hspace{0.2em}\frac{C_2 }{Q^4}\left(\!\ln\frac{Q^2}{\Lambda_{\QCD}^2}\!\right)^{\!-d_2}\hspace{-1.5em}\,, \ee
where $\Lambda_{\QCD}$ is the renormalization group invariant scale of QCD, while $d_2$ is related to the anomalous dimension 
of the operator $O_{\alpha\beta}^{f_2}$ and depends on the number of flavours:
\be\label{eq:d2}d_2= \frac{32}{3(33-2 N_f)}\,.\ee
The formula \eqref{eq:c2f} holds provided $Q$ is large, not only compared to $\Lambda_{\QCD}$, but compared to all of the  quark masses. In the intermediate range where $Q$ is large compared to $m_s$, but not large enough to activate the degrees of freedom of the heavy quarks, it should hold approximately, with $N_f \approx 3$. 

Since the perturbation series of $c_3^f(-Q^2)$ only starts at order $g^2$, the  asymptotic behaviour is suppressed by a factor of $\ln Q^2/\Lambda_{\QCD}^2$:
\be\label{eq:c3f} \sum_f c_3^f(-Q^2)\langle O^{f_2}\rangle^{p-n}\hspace{0.3em}\rightarrow\hspace{-1.9em}\rule[-0.6em]{0em}{0em}_{Q\rightarrow \infty}\hspace{0.2em}\frac{C_3 }{Q^4}\left(\!\ln\frac{Q^2}{\Lambda_{\QCD}^2}\!\right)^{\!-1-d_2}\hspace{-1.5em}\,. \ee

The scalar operator $\bar{f}f$ is of anomalous dimension as well, but the same is true of $m_f$ and the anomalies cancel: the  operator 
$m_f \bar{f}f$ is renormalization group invariant. This implies that, in the Wilson coefficient $c_1^f(-Q^2)$, the correction of order $g^2$ does not pick up a logarithmic enhancement if $Q^2$ becomes large and there is nothing to be summed up:
\be\label{eq:c1f} \sum_f c_1^f(-Q^2)\langle O^{f_0}\rangle^{p-n}\hspace{0.3em}\rightarrow\hspace{-1.9em}\rule[-0.6em]{0em}{0em}_{Q\rightarrow \infty}\hspace{0.2em}\frac{C }{Q^4} \,. \ee

Note that the  above relations only account for the leading logarithms. The perturbation series of the coefficient $c_1^f(-Q^2)$ does contain contributions of order $g^2$ that are not enhanced by a logarithm -- their role in the context of the Cottingham formula will be discussed in section \ref{sec:Subleading divergence}. 

Inserting the asymptotic expressions for the Wilson coefficients in  equations \eqref{eq:T2as} and \eqref{eq:Tbaras}, we obtain:
\bea\label{eq:Tas}
 \al\al \bar{T}(\nu,-Q^2)\hspace{0.3em}\rightarrow\hspace{-1.9em}\rule[-0.6em]{0em}{0em}_{Q\rightarrow \infty}\hspace{0.2em}
\frac{C}{Q^4} +\frac{C_3(\mbox{$\frac{1}{4}$}Q^2+\nu^2)}{Q^6}\left(\!\ln\frac{Q^2}{\Lambda_{\QCD}^2}\!\right)^{\!-1-d_2}\hspace{-2.5em},\no
 \al\al T_2(\nu,-Q^2)\hspace{0.3em}\rightarrow\hspace{-1.9em}\rule[-0.6em]{0em}{0em}_{Q\rightarrow \infty}\hspace{0.2em}
\frac{C_2}{Q^4}\left(\!\ln\frac{Q^2}{\Lambda_{\QCD}^2}\!\right)^{\!-d_2}\hspace{-1.5em}.\eea
While the coefficient $C$ is determined by the spin averaged matrix elements of a renormalization group invariant operator, 
\be C =\sumf\, Q_f^2\,\langle m_f\bar{f}f\rangle^{p-n}\,,\ee
$C_2$ and $C_3$ are related to the matrix elements of the spin 2 operator $O_{\alpha\beta}^{f_2}$,  which do depend on the renormalization convention used.

\section{Moments of the structure functions}
\label{sec:Moments}
Let us now compare the dispersive representation \eqref{eq:DRx} with the asymptotic formulae \eqref{eq:Tas} obtained from perturbation theory. Since the form factors rapidly tend to zero when $Q^2$ becomes large, the elastic part of the amplitudes does not show up in the asymptotic behaviour. The dispersion integrals approach moments of the structure functions:
\bea\label{eq:moments} &&M_k(Q^2)=\int_0^{\xth}\hspace{-0.5em}dx F_k(x,Q^2)\,,\hspace{1em}k=2,L\\
&&\bar{M}(Q^2)=\int_0^{\xth}\hspace{-0.5em}dx \bar{F}(x,Q^2)\,.\nonumber\eea
In our decomposition of the dispersive representation, the asymptotic behaviour \eqref{eq:Tas} boils down to a set of  conditions on the subtraction function and on the lowest moments of $\bar{F}$, $F_2$ and $F_L$: 
\bea \label{eq:Sas}\bar{S}(-Q^2)\al\hspace{1em}\rightarrow\hspace{-1.9em}\rule[-0.6em]{0em}{0em}_{Q^2\rightarrow \infty}\hspace{0.2em}\al \frac{C}{Q^4}\,,\\
\bar{M}(Q^2)\al\hspace{1em}\rightarrow\hspace{-1.9em}\rule[-0.6em]{0em}{0em}_{Q^2\rightarrow \infty}\hspace{0.2em}\al\frac{C_3}{4m^2}\left(\!\ln\frac{Q^2}{\Lambda_{\QCD}^2}\!\right)^{\!-1-d_2}\hspace{-1.5em},\label{eq:Mbar}\\
M_2(Q^2)\al\hspace{1em}\rightarrow\hspace{-1.9em}\rule[-0.6em]{0em}{0em}_{Q^2\rightarrow \infty}\hspace{0.2em}\al\frac{C_2}{4m^2}\left(\!\ln\frac{Q^2}{\Lambda_{\QCD}^2}\!\right)^{\!-d_2}\hspace{-1.5em},\label{eq:M2}\\
M_L(Q^2)\al \hspace{1em}\rightarrow\hspace{-1.9em}\rule[-0.6em]{0em}{0em}_{Q^2\rightarrow \infty}\hspace{0.2em}\al 
\frac{C_3}{4m^2}\left(\!\ln\frac{Q^2}{\Lambda_{\QCD}^2}\!\right)^{\!-1-d_2}\hspace{-2.5em}.\label{eq:ML}\eea

In the literature, the perturbative predictions for the moments have been compared in detail with experi\-ment \cite{Buras}. The parametrization we will be using for the structure functions is based on the DGLAP equations \cite{DGLAP,DGLAP1,DGLAP2}. These ensure that the behaviour of the Compton amplitude in the deep inelastic region is consistent with perturbation theory.  

\section{Prediction for the constant C}
\label{sec:C}
Neglecting isospin breaking effects of second order, the neutron matrix elements of $e^2\dbar d$ agree with the proton matrix elements of $e^2\ubar u$ and vice versa. The constant $C$ can thus be expressed  in terms of proton matrix elements:
\be\label{eq:C} C= \frac{4m_u-m_d}{9}\langle p|\bar{u}u-\bar{d}d|p\rangle\,.\ee

The matrix element of the operator $\bar{u}u-\bar{d}d$ also determines the leading contribution to the QCD part of the proton-neutron mass difference (see e.g.~\cite{GL1982}):
\be\label{eq:mpnQCD}m_{\QCD}=\frac{m_u-m_d}{2m}\langle p|\bar{u}u-\bar{d}d|p\rangle\left\{1+O(m_u-m_d)\right\}\,.\ee
This shows that the constant $C$  is  related to the value of the proton-neutron mass difference in the absence of the e.m.~interaction:
\be C= \frac{2(4m_u-m_d)}{9(m_u-m_d)}\,m\, m_{\QCD} +O(m_u-m_d)\,.
\label{eq:CmQED}\ee

Once we have determined the e.m.~part, the observed mass difference will provide us with a value of $m_{\QCD}$ and hence also with a value of the constant $C$. 

Actually, however, the precise value of  $C$ is not crucial in our context. For our purpose, the crude estimate $m_{\QCD}\approx -2\MeV$ is good enough. The quark mass ratio $r=(4m_u-m_d)/(m_d-m_u)$ is determined by $m_u/m_d$, but is not yet known very firmly. The FLAG result $m_u/m_d=0.513(39)$ (for $N_f=2+1+1$) \cite{FLAG} implies $r=2.16(42)$. There is a totally independent determination of the mass ratio $Q$, based a low energy theorem for the decay $\eta\rightarrow3\pi$ \cite{GL1985,Kambor1996,Anisovich1996}. A recent analysis of the data on this basis leads to $Q=22.1(7)$ \cite{CLLP}. Combining this result with the well-determined ratio $m_s/m_{ud}=27.23(10)$ \cite{FLAG}, we obtain $m_u/m_d=0.450(25)$ and $r=1.46(25)$. As pointed out in \cite{CLLP}, the origin of the difference could be identified by calculating the corrections to the low energy theorem on the lattice, but this yet needs to be done. The outcome for the constant $C$ is tiny. With the value $r=1.46$, we obtain:
\be\label{eq:Capprox} C\approx 6\cdot 10^{-4} {\GeV}^2\,.\ee
The reason why the value turns out to be so small is that $C$ vanishes  in the chiral limit. It implies that $C$ is small compared to $C_2$ and $C_3$. Accordingly, it takes very large values of $Q^2$ for the singularities generated by the operators of spin 0 to finally dominate over those associated with operators of spin 2.

\section{Renormalization}
\label{sec:Renormalization}
The asymptotic behaviour of the subtraction function in equation \eqref{eq:Sas} implies that the corresponding contribution to $m_{\QED}$ is logarithmically divergent. The leading  divergence is determined by the coefficient $C$:
\be \label{eq:mlog} \int_0^{\Lambda^2}\hspace{-1em} dQ^2 Q^2 \bar{S}(-Q^2)\hspace{0.5em}\rightarrow\hspace{-1.9em}\rule[-0.6em]{0em}{0em}_{\Lambda\rightarrow \infty}\hspace{0.2em} C \ln\Lambda^2\,.\ee
A logarithm also occurs in the electromagnetic renormalization of the  bare QCD coupling constant and of the bare quark masses (see e.g.~\cite{GL1982}):
\bea\label{eq:Deltag}\al\al\Delta g= -\frac{e^2g^3}{256\pi^4}\sumf Q_f^2 \ln\frac{\Lambda^2}{\mu^2}\,,\\
\label{eq:Deltam}\al\al\Delta m_f=\frac{3e^2}{16\pi^2}Q_f^2m_f\ln\frac{\Lambda^2}{\mu^2}\,. \eea
The scale $\mu$ of the logarithm is a matter of convention --  picking a value for $\mu$ amounts to fixing the ambiguity in the decomposition \eqref{eq:mpn} of the mass difference into contributions arising from the e.m.~and strong interactions, respectively. 

In the difference between the masses of proton and neutron, the e.m.~renormalization of the coupling constant only yields a contribution of second order in isospin breaking -- we are neglecting such effects. The renormalization of the quark masses, on the other hand, does not drop out in the difference. In the Lagrangian, the corresponding counter term reads
\be\label{eq:DeltaL}\Delta\mathcal{L}=\sumf \Delta m_f\fbar f\,.\ee
The corresponding shift in the mass of a particle is given by $-\langle p|\Delta\mathcal{L}|p\rangle/2m$. Accordingly, the change in the proton mass generated by the renormalization of $m_u$ and $m_d$ is given by
\be \Delta m^p =- \frac{1}{2m^p}\{\langle p|\ubar u|p\rangle\Delta m_u+\langle p|\dbar d|p\rangle\Delta m_d\}\,. \ee
Again neglecting effects of second order in  isospin breaking, the neutron matrix elements can be expressed in terms of those of the proton:
\be \Delta m^n =- \frac{1}{2m^n} \{\langle p|\dbar d|p\rangle\Delta m_u+\langle p|\ubar u|p\rangle\Delta m_d\}\,. \ee
Collecting terms  and neglecting second order isospin breaking effects, we obtain the following expression for the counter term $\Delta m^\Lambda=\Delta m^p-\Delta m^n$:
\be \Delta m^\Lambda=-  \frac{\alphaem}{24\pi m}(4m_u-m_d)\langle p|\ubar u-\dbar  d|p\rangle \ln\frac{\Lambda^2}{\mu^2} \,.\ee 
Comparison with the expression \eqref{eq:C} for the constant $C$ that determines the asymptotic behaviour of the subtraction function shows that the two quantities are related by
\be \Delta m^\Lambda=-N C\ln\frac{\Lambda^2}{\mu^2}\,,\ee
where the normalization factor $N$ is specified in equation \eqref{eq:N}. As it should be, the logarithm in the integral \eqref{eq:mlog} over the subtraction function thus cancels the one  in the renormalization \eqref{eq:Deltam} of the quark masses: the leading divergences occurring in the expression \eqref{eq:mS} for $m_{\bar{S}}$ drop out. 

\section{Subleading divergence}\label{sec:Subleading divergence}
As mentioned above, the asymptotic  formulae \eqref{eq:Tas} only account for the leading logarithms -- they are valid only up to corrections of order $g^2$. This applies, in particular also to the Wilson coefficient of the spin 0 operator that is responsible for the logarithmic divergence of the Cottingham formula. The correction of order $g^2$ gives rise to a theoretical issue, which does not appear to be covered in the literature and which we briefly wish to address.  

When $Q^2$ becomes large, the effective strength of the interaction decreases in proportion to $1/\ln Q^2$. Those corrections of order $g^2$ in the Wilson coefficients or in the counter term $\Delta m^\Lambda$ that do not pick up a logarithmic enhancement are asymptotically small compared to the leading terms. This does not ensure, however, that the corresponding contributions to $m_{\bar{S}}$ remain finite when the cutoff is removed: the corresponding contributions instead grow in proportion to
\be \int^{\Lambda^2}\hspace{-1em}dQ^2\;\frac{1}{Q^2\ln Q^2}= \ln\ln \Lambda^2\,.\ee 

The coefficient of order $e^2g^2$  in the renormalization of the quark masses in QCD+QED is known \cite{Chetyrkin,Grozin}:
\bea\label{eq:RGm} \mu\frac{d m_f}{d\mu} \al=\al-\gamma_m(g,e) m_f\,,\\
 \gamma_m(g,e) \al=\al\gamma_0\frac{g^2}{4\pi^2}+\gamma_1\frac{g^4}{16\pi^4}+O(g^6)\label{eq:gammam}\\
\al+\al\frac{3e^2Q_f^2}{8\pi^2}\{1+\frac{g^2}{12\pi^2}+O(g^4)\}+O(e^4)\,,\nonumber\eea
where $\gamma_0=2$ and $\gamma_1=\frac{101}{12}-\frac{5}{18}N_f$ are the well-known coefficients relevant for mass renormalization in QCD. The counter term $\Delta m^\Lambda$ considered above is related to the term of order $e^2$ in equation \eqref{eq:gammam}. 

On the other hand, the contributions of order $g^2$ in the Wilson coefficient were considered by Shifman, Vainshtein and Zakharov, more than 40 years ago \cite{SVZ}.  Equations (4.15) and (4.18) in this reference indicate that, in the notation used above, the coefficient $C$ picks up the same correction as the counter term, 
\be\label{eq:Cg}C\rightarrow C\left\{1+\frac{g^2}{12\pi^2}+O(g^4)\right\}\,.\ee
 As this amounts to combining results obtained within two different regularization schemes (cutoff in euclidean momentum space, dimensional regularization) it must be taken with a grain of salt, but it does indicate that the divergences of the type $\ln\ln\Lambda^2$ cancel. Irrespective of the regularization used, the renormalization of coupling constants and quark masses must remove the divergences also at the subleading level.
 
Numerically, the perturbative corrections are not important, because, as pointed out above, chiral symmetry suppresses the entire contribution to the mass difference from the region where perturbation theory applies. In that region, the corrections are even smaller than the leading terms -- they are in the noise of our calculation and we neglect them. The limit $\Lambda\rightarrow\infty$ in formula \eqref{eq:mS} can then be done explicitly. The result can be written in the form
\be\label{eq:mSbar}m_{\bar{S}}=N\int_0^\infty \hspace{-0.8em}dQ^2 Q^2 \left\{\bar{S}(-Q^2)-\frac{C}{(\bar{\mu}^2+Q^2)^2}\right\}\,,\ee
with $\bar{\mu}=\mu \exp(-\frac{1}{2})$.

\section{Input used for the structure functions}
\label{sec:Input}
For the numerical evaluation of the inelastic contributions, we need a representation for the difference between the structure functions of proton and neutron, and not only for the relatively well explored quantity $F_2$, but also for the longitudinal component $F_L$, which is known less well. At low values of $Q^2$, we closely follow the analysis of \cite{GHLR} and distinguish three different regions in the centre of mass energy $W=\sqrt{s}$ (numerical values for $W$ and $Q^2$ are given in GeV units): 

\begin{description}
\item[(i)] For the range $W<1.3$, we rely on the parametri\-zations of the structure functions of MAID and DMT \cite{Drechsel:2007if,
 Kamalov:2000en,Hilt:2013fda}
-- we refer to these as MD. Both of them are accessible on the MAID home page \cite{MAID}. We identify the central values of the structure functions in this region with the mean of the two parametrizations and use the difference as an error estimate (half of the difference would suffice to cover the two). The green error band in Fig.~\ref{fig:Fbar}, shows the corresponding representation of the structure function $\bar{F}$ for $Q^2=1$.
\begin{figure}
\resizebox{0.48\textwidth}{!}{%
\includegraphics{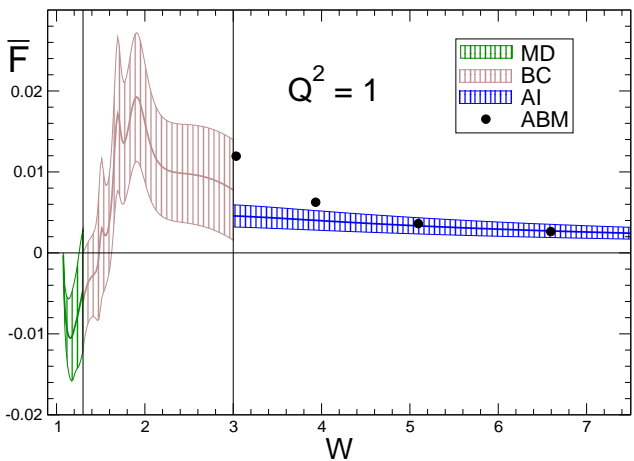}
}
\caption{Structure function $\bar{F}$ versus $W$ at $Q^2=1$ (GeV units for $Q^2$ and $W$). For $W<1.3$ and $1.3<W<3$, we use the representations labeled MD \cite{Drechsel:2007if, Kamalov:2000en,Hilt:2013fda,MAID} 
 and BC \cite{Bosted1,Bosted2}, respectively. In the region $W>3$ we rely on two different parametrizations: the Regge representation AI \cite{GVMD}  and the ABM table. For further explanations, see text.\label{fig:Fbar}}       \end{figure}
 
\item[(ii)] In the interval $1.3< W < 3 $, we make use of the representation due to Bosted and Christy (BC) \cite{Bosted1,Bosted2}. It contains a wealth of information, but suffers from several shortcomings that are discussed in detail in section 5.1 of \cite{GHLR}. Part of
the problem originates in the fact that the longitudinal cross section is more difficult to measure than the transverse one. In \cite{Bosted1,Bosted2} it is assumed that the ratio $R=\sigma_L/\sigma_T$ of the neutron cross sections is the same as for the proton. In the region where the Pomeron dominates, this holds to good accuracy, but we need the difference between the two, where Pomeron exchange drops out. The assumption amounts to using an approximation and introduces a systematic error that is not easy to estimate. 

In our opinion, the procedure used in \cite{GHLR} to cope with the uncertainties in the region $1.3< W < 3 $ is on the conservative side and we adopt it in the present work: we treat the transverse and longitudinal cross sections as independent and assign an uncertainty in $\sigma_T^{p-n}$ and $\sigma_L^{p-n}$ of $8\%$ of $\sigma_T^p$ and $8\%$ of $\sigma_L^p$, respectively. In part of phase space, this may well overestimate the uncertainties considerably -- a reanalysis of the data in the resonance region would be most welcome. The structure of the brown error band reflects the resonances occurring in this region.
 
\item[(iii)] For $W> 3,\,Q^2<1$, we rely on the parametrization of the proton structure functions due to Alwall and Ingelman (AI) \cite{GVMD}. It represents the amplitude as a sum of a contribution from the Pomeron and one from the $a_2$. In the difference between the proton and neutron amplitudes, the Pomeron drops out. We assume that the couplings of the $a_2$ to proton and neutron are approximately SU(3)-symmetric and attach an uncertainty of $30\%$ to the representation for the difference between proton and neutron obtained on this basis. The blue band in Fig.~\ref{fig:Fbar} shows the corresponding uncertainty range at $Q^2=1$. For details, we refer to section 5.1 in \cite{GHLR}.

\item[(iv)] In the region $W>3$, $Q^2>1$, we use the solution of the DGLAP equations constructed by Alekhin, Bl\"umlein and Moch, who obained numerical values for the structure functions over a wide range: $1 \leq Q^2\leq2\cdot 10^5 $ and $10^{-7}\leq x\leq 0.99$. The values of $F_2(x,Q^2)$ and $F_L(x,Q^2)$ are listed for the proton as well as for the neutron on a grid of $60\times 98$ points. We thank Johannes Bl\"umlein for providing us with this table, which we refer to with the acronym ABM. The underlying analysis is described in \cite{ABM1,ABM2,ABM3}. In the region where we make use of these results (see below), we estimate the uncertainty in the values obtained from ABM for $\bar{F}_{p-n}$ and $F_2^{p-n}$  at 30\%.

\vspace{2mm}
In the deep inelastic region, asymptotic freedom leads to very strong constraints, particularly for the structure function $F_L$. The strength of these constraints is clearly visible at leading order of the perturbative expansion, where $F_L$ is given by an integral over $F_2$. The DGLAP equations extend this relationship to higher orders, by means of the renormalization group. In our framework, the properties of $F_L$ play a crucial role in the evaluation of the sum rule for the subtraction function. The theoretical constraints on this quantity are very important for our analysis, particularly also because the raw experimental information for $F_L$ is much weaker than the one for $F_2$. 
\end{description}

The black dots in Fig.~\ref{fig:Fbar} show the numbers obtained for $W$ and $\bar{F}$ from the entries for $x$, $F_2$ and $F_L$, at the lowest value of $Q^2$ listed in the ABM table, $Q^2=1$, and $W>3$.  While the result agrees very well with AI for $W>5$, the two representations do differ at lower values of $W$. Since the DGLAP equations rely on perturbation theory, we should not be surprised to find deviations at low momenta, i.e.~in the region where $Q^2$ as well as $W$ are small.  
\section{\texorpdfstring{Polarizabilities, $\bar{S}(0)$}{Polarizabilities \textoverline{S}(0)}}
\label{sec:Polarizabilities}
Two low energy theorems relate the values of $T_1$ and $T_2$ at $q=0$ to the polarizabilities of proton and neutron:\footnote{For a derivation see e.g.~appendix B.8 in \cite{GHLR}).}
\bea T_1\al =\al T_1^{\el} -\frac{\kappa^2}{4m^2}-\frac{m}{\alphaem}\beta_M+O(\nu^2,q^2)\,.\\
T_2 \al =\al T_2^{\el} +\frac{m}{\alphaem}(\alpha_E+\beta_M)+O(\nu^2,q^2)\,,\nonumber\eea
where $\kappa$ is the anomalous magnetic moment of the particle (these relations hold separately for proton and neutron). The dispersion relation for $T_2$ converts the second one into the Baldin sum rule \cite{Baldin}, which expresses the sum $\alpha_E+\beta_M$ as an integral over the cross section for photoproduction. For the combination $\bar{T}=T_1+\frac{1}{2}T_2$ we are working with, the low energy theorem involves the difference between the electric and magnetic polarizabilities: 
\be\label{eq:TbarLET}\bar{T}=\bar{T}^{\el} -\frac{\kappa^2}{4m^2}+\frac{m}{2\alphaem}(\alpha_E-\beta_M)+O(\nu^2,q^2)\,.\ee
It fixes the value of the subtraction function $\bar{S}(q^2)$ at $q^2=0$ in terms of the polarizabilities: 
\be \label{eq:Sbar0alphabeta}\bar{S}(0)=-\frac{1}{4m^2}\kappa^2+\frac{m}{2\alphaem}(\alpha_E-\beta_M)\,,\ee
For $q^2=0$, our sum rule for the subtraction function thus represents an analog of the Baldin sum rule: it determines the value of the difference between the electric and magnetic polarizabilities rather than their sum, in terms of the structure functions. While the Baldin sum rule directly follows from the unsubtracted dispersion relation for $T_2$, the one for $\alpha_E-\beta_M$ relies on Reggeon dominance. 

The integrals over the structure functions relevant for the evaluation of the subtraction function in the dispersion relation for $T_1$ at small values of $Q^2$ were analyzed in detail in section 5 of \cite{GHLR}. As shown there, the prediction for the electric polarizability comes with comparatively small uncertainties:\footnote{As usual, the polarizabilities are given in units of $10^{-4}\, \text{fm}^3$.} 
\be\label{eq:alphaGHLR} \alpha_E^{p-n}=-1.7(4)\,\mbox{\cite{GHLR}}\,.\ee  
The averages for proton and neutron quoted by the Particle Data Group yield $\alpha_E^{p-n}=-0.6(1.2)$  \cite{PDG}. The fact that experiment agrees with the prediction within errors provides a test of Reggeon dominance. 

For a review of the currently available information about the polarizabilities, in particular also of the analysis based on chiral effective theories, we refer to \cite{Grieshammer:2019nlx,Lensky,Hagelstein}. In the framework of  $\chi$PT, the representation of the virtual Compton scattering amplitude has been worked out to first nonleading order \cite{AHLP}. In this reference, the low energy singularity generated by the $\Delta(1232)$ resonance is explicitly accounted for. It will be of considerable interest to compare the result of this analysis for the slope of the subtraction function at $Q^2=0$ with the solution of the sum rule that follows from Reggeon dominance constructed in the present paper. 

The Baldin sum rule and the data on photoproduction  imply that the sum $\alpha_E+\beta_M$ is known more accurately than the individual terms. For this reason, it is useful to treat $\alpha_E\pm\beta_M$ as the two independent quantities.  The results quoted for proton and neutron in the compilation of Melendez et al.~lead to
\bea\label{eq:Melendez1}
(\alpha_E+\beta_M)^{p-n}\al=\al-1.20(45)\,\text{\cite{Melendez}}\,,\\
\label{eq:Melendez2}(\alpha_E-\beta_M)^{p-n}\al=\al-0.4(3.1)\,\text{\cite{Melendez}}\,.\eea
Combining the prediction \eqref{eq:alphaGHLR}, which is based on Reggeon dominance, with the result \eqref{eq:Melendez1} obtained from photoproduction, we obtain a prediction for the magnetic polarizability, which is slightly more accurate than the one given in \cite{GHLR}:
\be\label{eq:betaGHLRM} \beta_M^{p-n}=0.5(6)\,.\ee

There were early attempts at calculating the electric polarizabilities on the lattice \cite{Detmold:2006vu,Detmold:2009dx,Detmold:2010ts}, based on turning on a constant external electric field, but they did not reach a level where the results could be compared with the experimental determinations in a meaningful way. The very recent lattice determination of the magnetic polarizabilities, however, which makes use of a constant external magnetic field, does yield a remarkably precise value for $\beta_M^{p-n}$,  
\be\label{eq:betaB}\beta_M^{p-n}=0.80(28)(4)\,\text{\cite{Bignell}}\,,\ee
in good agreement with our predicton in equation \eqref{eq:betaGHLRM}.

In connection with the proton-neutron mass difference, the polarizabilities are of interest because they determine the value of the subtraction function $\bar{S}(q^2)$ at $q^2=0$, according to \eqref{eq:Sbar0alphabeta}. The prediction \eqref{eq:alphaGHLR} for $\alpha_E^{p-n}$ and the experimental value \eqref{eq:Melendez1} of $(\alpha_E+\beta_M)^{p-n}$ imply
\be\label{eq:Sbar0predM}\bar{S}(0)=-1.71(77)\GeV^{-2}\,.\ee
The uncertainty is significantly smaller than the one obtained from the experimental value  
of $(\alpha_E-\beta_M)^{p-n}$:
\be\label{eq:Sbar0M}\bar{S}(0)=-0.2(2.6)\GeV^{-2} \text{\cite{Melendez}}\,.\ee
On the other hand, combining the lattice result \eqref{eq:betaB} for the magnetic polarizability with the experimental value \eqref{eq:Melendez1} of $(\alpha_E+\beta_M)^{p-n}$, we obtain a result for the subtraction function at the origin that is even slightly more precise than the prediction:
\be\label{eq:Sbar0B}\bar{S}(0)=-2.22(60)\GeV^{-2}\,.\ee
The fact that, within errors, this result agrees with the prediction \eqref{eq:Sbar0predM} amounts to a more stringent test of the Reggeon dominance hypothesis than the one discussed above. It is important to pursue the determination of the polarizabilities; in particular, the pioneering lattice result which provides such a test calls for confirmation.

\section{\texorpdfstring{Subtraction function at low $Q^2$}{Subtraction function at low Q \textasciicircum 2}}
\label{sec:Low}
\begin{figure}
\resizebox{0.48\textwidth}{!}{%
\includegraphics{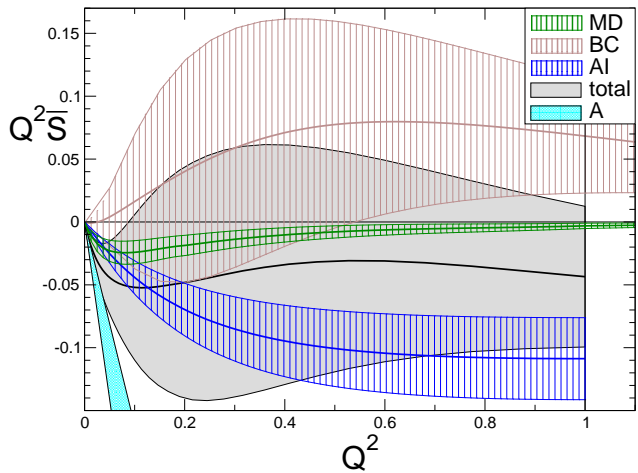}
}
\caption{Subtraction function at low values of $Q^2$ (GeV units, $Q^2\bar{S}$ is dimensionless). The black line and the gray region depict central value and error band attached to our result for $Q^2\leq 1$. It represents the sum of the contributions from the regions  $W\leq 1.3$ (MD),  $1.3\leq W\leq 3$ (BC) and $3\leq W$ (AI), which are discussed in the text. This part of our representation for the subtraction function stops at $Q^2=1$ because it relies on a Regge representation that is not valid beyond this point. The cyan-coloured wedge labeled A represents the tangent at $Q^2=0$ obtained with the magnetic polarizability of \cite{Bignell}, see equation \eqref{eq:Sbar0B}. \label{fig:Q2SbarLow}}       \end{figure}

Next, we discuss the solution of the sum rule \eqref{eq:Sum rule} for $Q^2<1$, where the parametrizations listed in (i) -- (iii) suffice. 
Fig.~\ref{fig:Q2SbarLow} displays the contributions arising from the various regions of phase space. 

The interval of integration in \eqref{eq:Sum rule} is split into three parts that correspond to the regions where we are using the representations MD, BC and AI, respectively. The values of $x$ where $W=1.3$ and $W=3$ are denoted by $x_a$ and $x_b$, respectively. In the first two parts, the integration over the term $\bar{F}^R/x^2$ can explicitly be done -- we book these contributions together with the term involving the Reggeon residues in $\bar{S}_{AI}$. For $Q^2<1$, the solution of the sum rule then takes the form
\bea\label{eq:SMDBCAI}\bar{S}\al=\al\bar{S}_{\mbox{\tiny MD}}+\bar{S}_{\mbox{\tiny BC}}+
\bar{S}_{\text{\tiny AI}}\,,\\
\bar{S}_{\text{\tiny MD}}\al=\al\int_{x_a}^{\xth}\hspace{-1.3em}dx\,\frac{\bar{F}(x,Q^2)}{x^2(Q^2+m^2 x^2)}\,,\no
\bar{S}_{\text{\tiny BC}}\al=\al\int_{x_b}^{x_a}\hspace{-1.3em}dx\,\frac{\bar{F}(x,Q^2)}{x^2(Q^2+m^2 x^2)}\,,\no
Q^2\bar{S}_{\text{\tiny AI}}\al=\al\int_0^{x_b}\hspace{-1.3em}dx\,\frac{\bar{F}(x,Q^2)-\bar{F}^{\indR}(x,Q^2)}{x^2}\no\al-\al \sum_{\alpha>0}\frac{b_\alpha(Q^2)}
{\alpha\, x_b^\alpha}-m^2\hspace{-0.4em} \int_0^{x_b}\hspace{-1.3em}dx\, \frac{\bar{F}(x,Q^2)}{Q^2+m^2 x^2}\,.\nonumber\eea
The term $\bar{S}_{\mbox{\tiny MD}}(-Q^2)$ includes the most prominent low energy phenomenon, the resonance $\Delta(1232)$. Isospin conservation ensures that the couplings of this state to proton and neutron are the same, so that the resonance does not show up at all in the subtraction function relevant for the difference between the two. Indeed, as shown by the green band, the contributions from this region are small.

In the region of the higher resonances, we rely on the BC representation of the structure functions. The brown error band indicates the price to pay with the error estimate specified in section \ref{sec:Input}: the largest uncertainty in our evaluation of the mass difference stems from there. 

The blue band depicts the function $S_{\text{\tiny AI}}$. Since the Regge representation AI we are using in this region is restricted to $Q^2< 1$, the band stops at $Q^2=1$. 

The plot shows that the contributions from MD and AI are negative, while the one from BC is predominantly positive. The net central value of $\bar{S}(-Q^2)$, which is indicated by the black line, is rather small and negative, but the uncertainty attached to it (gray error band) excludes positive values only in the vicinity of $Q^2=0$. 

The cyan-coloured wedge labeled A represents the tangent $Q^2\bar{S}(-Q^2)=Q^2 \bar{S}(0)+\ldots$ calculated with the value of $\bar{S}(0)$ in equation \eqref{eq:Sbar0B} (lattice result for magnetic polarizability plus Baldin sum rule). It confirms that at small values of $Q^2$, the subtraction function is negative.

\section{\texorpdfstring{Intermediate values of $Q^2$}{Intermediate values of Q \textasciicircum 2}}
\label{sec:Intermediate values}The representations of MD and BC are valid also for $Q^2>1$, but for AI, this is not the case. We instead rely on ABM. The formula for the corresponding contribution to the subtraction function is the same as for AI, but the representation for $\bar{F}$ consists of a numerical table rather than an algebraic parametrization like the one of AI. 

The leading term in equation \eqref{eq:FR} stems from the $a_2$ and has $\alpha\simeq 0.55$. In order to determine the corresponding coefficient $b_\alpha$, we focus on small values of $x$ and approximate the numbers  for $\bar{F}$ obtained from the ABM table at a given value of $Q^2$ with an approximation of the form 
\be\label{eq:FRpoly} \hat{F}=x^{1-\alpha}(b_\alpha +b_\alpha' x + b_\alpha'' x^2)\,.\ee 
The coefficients $b_\alpha,b_\alpha',b_\alpha''$ depend on $Q^2$. We determine them by minimizing the sum of the squares of the differences between the para\-me\-trization and the ABM values over a suitable interval. At very small values of $x$, the numerical noise in the entries of the table hides the signal while if $x$ is too large, the approximation used breaks down -- we find that $10^{-4}<x<x_1$ with $x_1=3 \cdot 10^{-2}$ represents a suitable range. In the grid of $x$-values used in the ABM table, this range contains points \# 15 to 25. We fix the parameter $b_\alpha''$ with continuity at point \# 24 and treat the coefficients $b_\alpha,b_\alpha'$ as free parameters. For a given value of $Q^2$, the minimization then fixes these. In particular, the procedure determines the Reggeon residue, which according to \eqref{eq:FR} is given by $\beta_\alpha=\frac{1}{2}Q^{-2(\alpha+1)}b_\alpha$.  

\begin{figure}{%
\resizebox{2.7cm}{!}{\includegraphics*{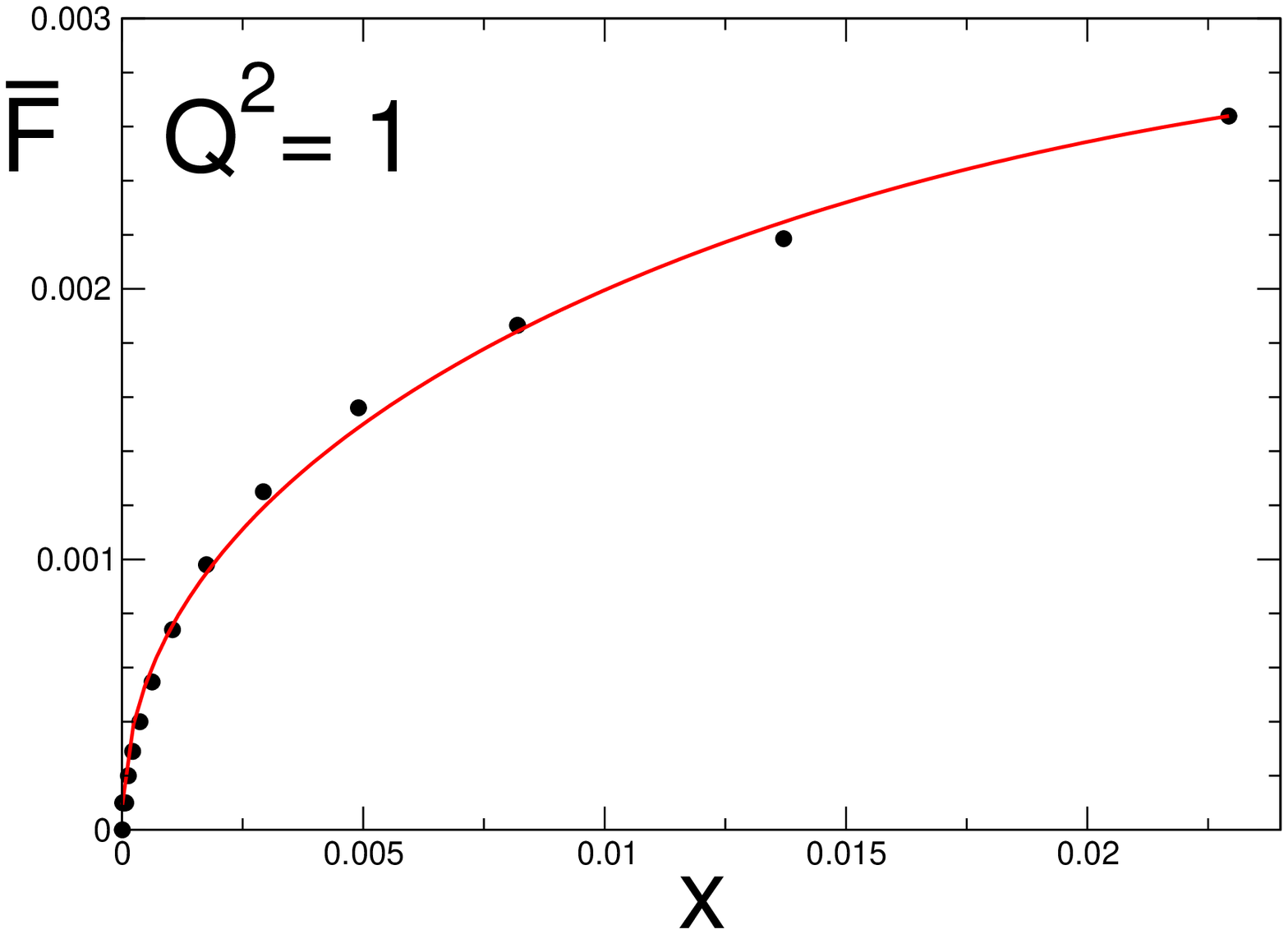}}   \resizebox{2.7cm}{!}{\includegraphics*{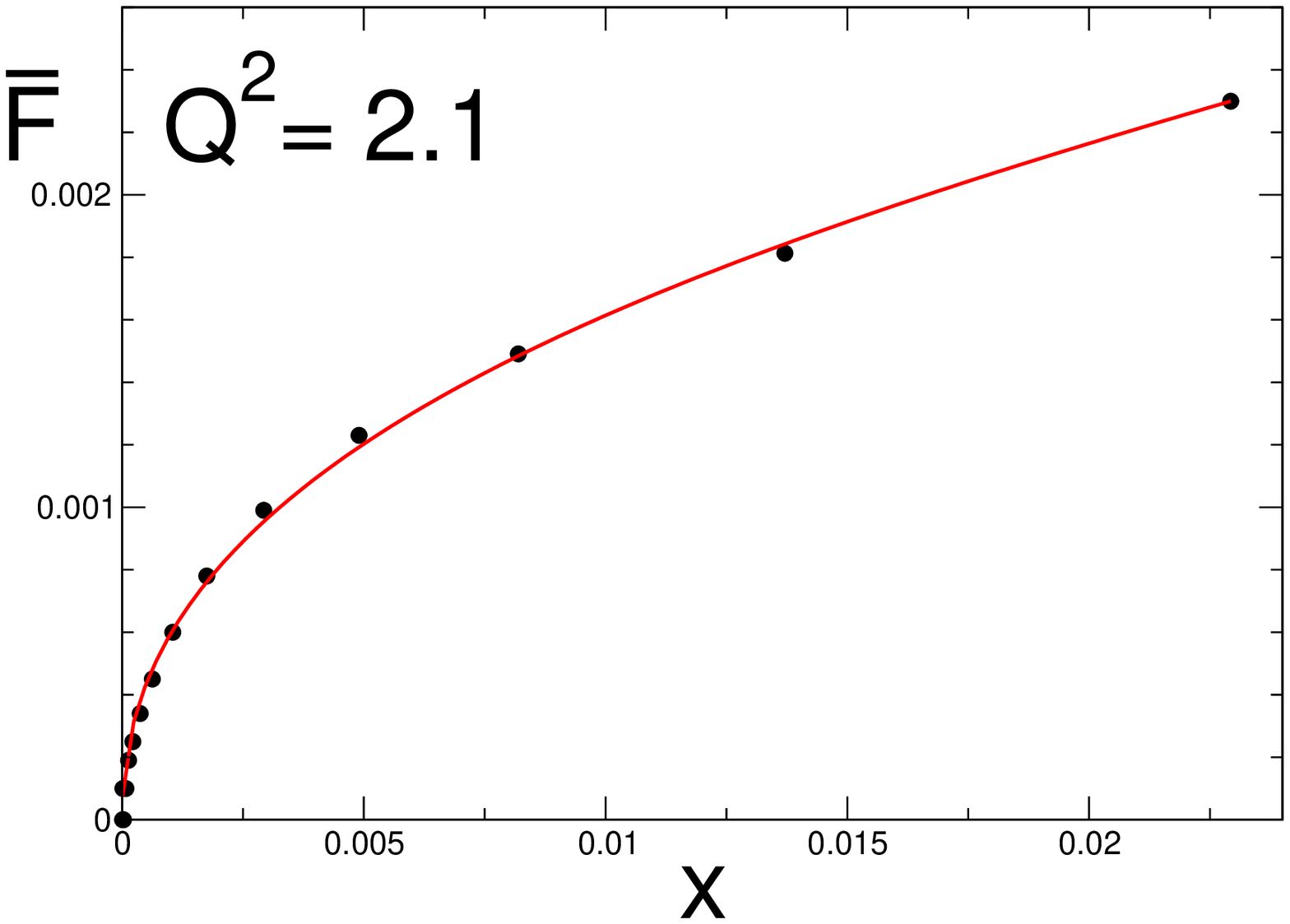}}
\resizebox{2.7cm}{!}{\includegraphics*{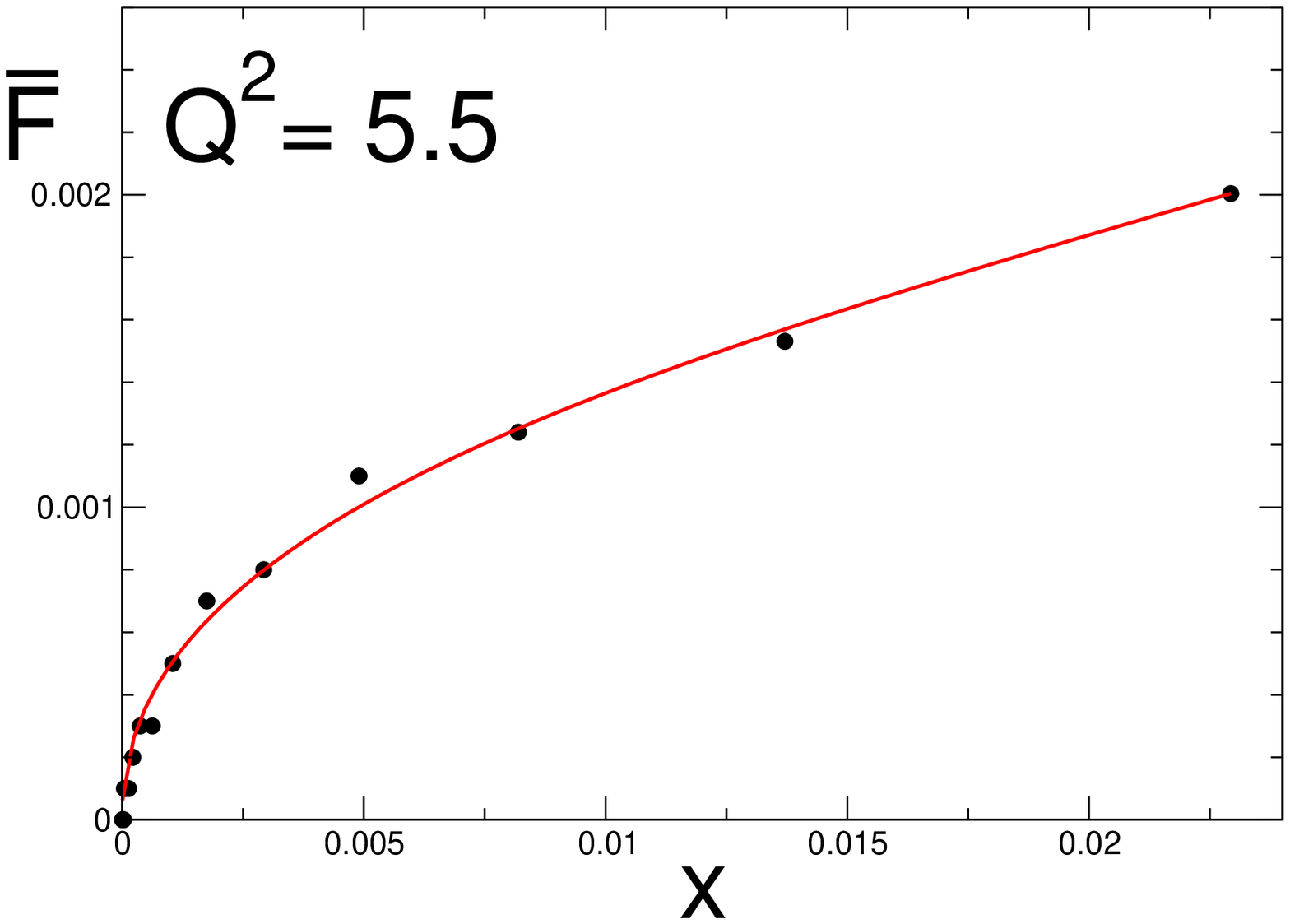}} \\ \resizebox{2.7cm}{!}{\includegraphics*{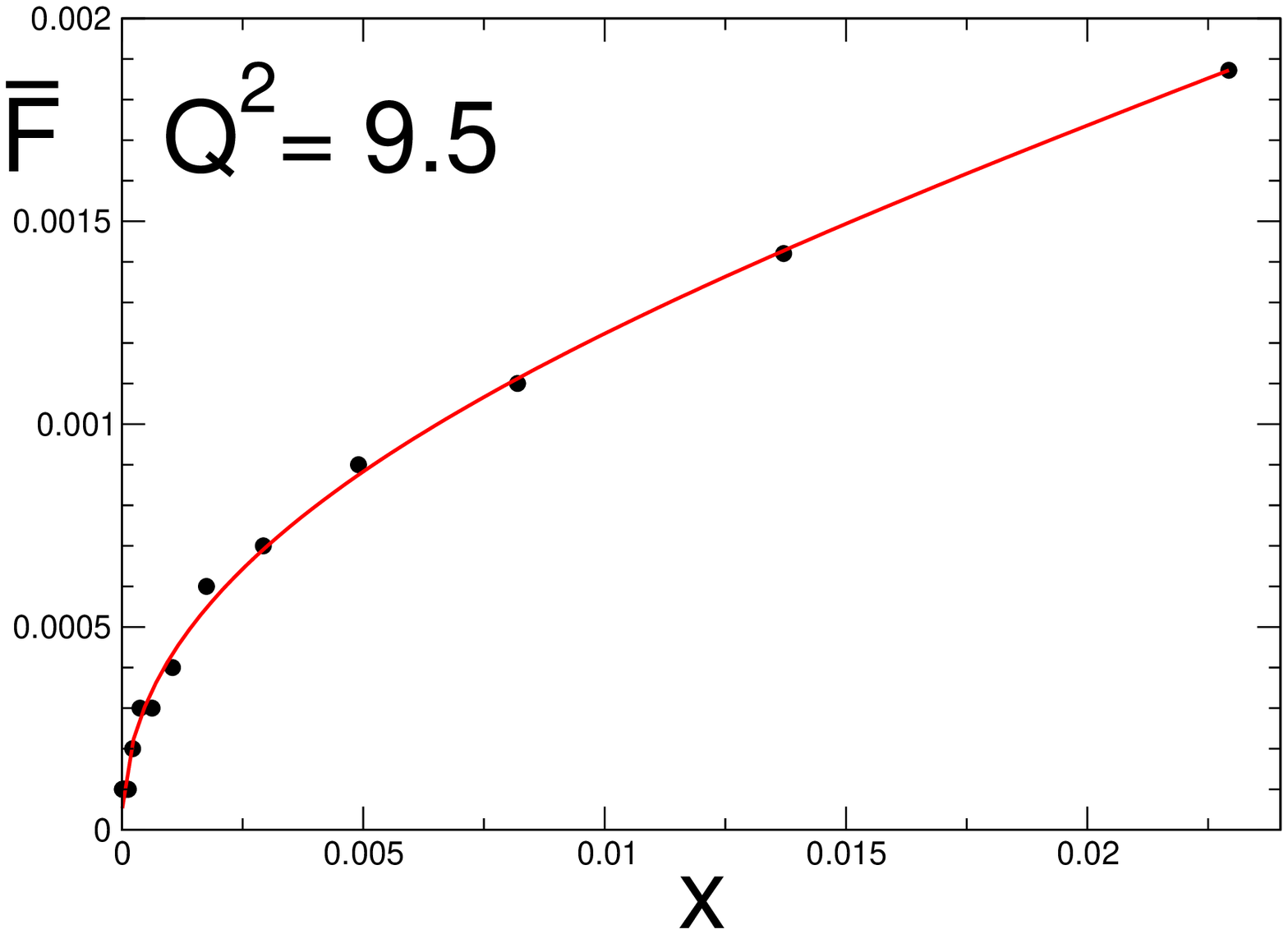}} 
\resizebox{2.7cm}{!}{\includegraphics*{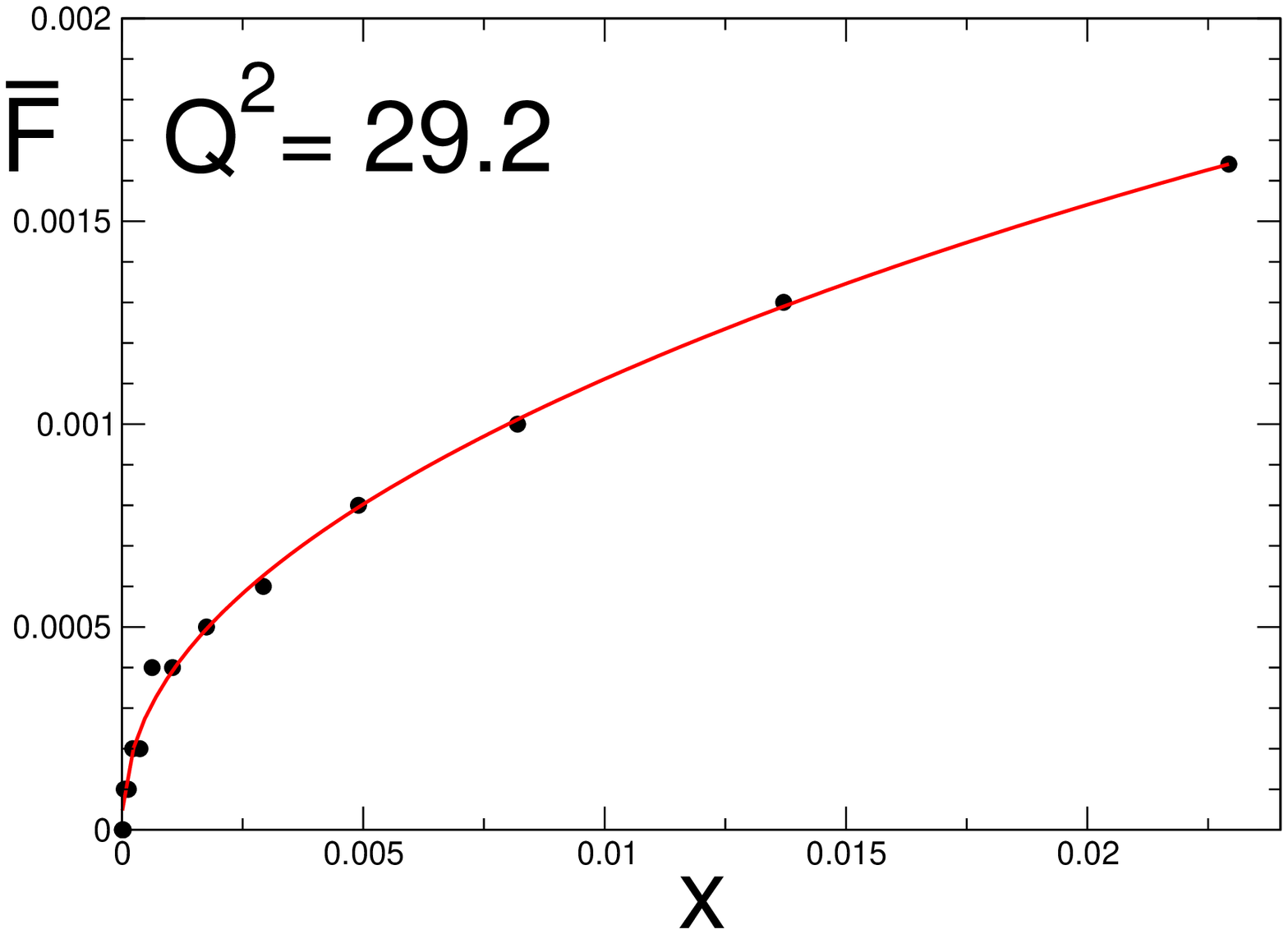}} \resizebox{2.7cm}{!}{\includegraphics*{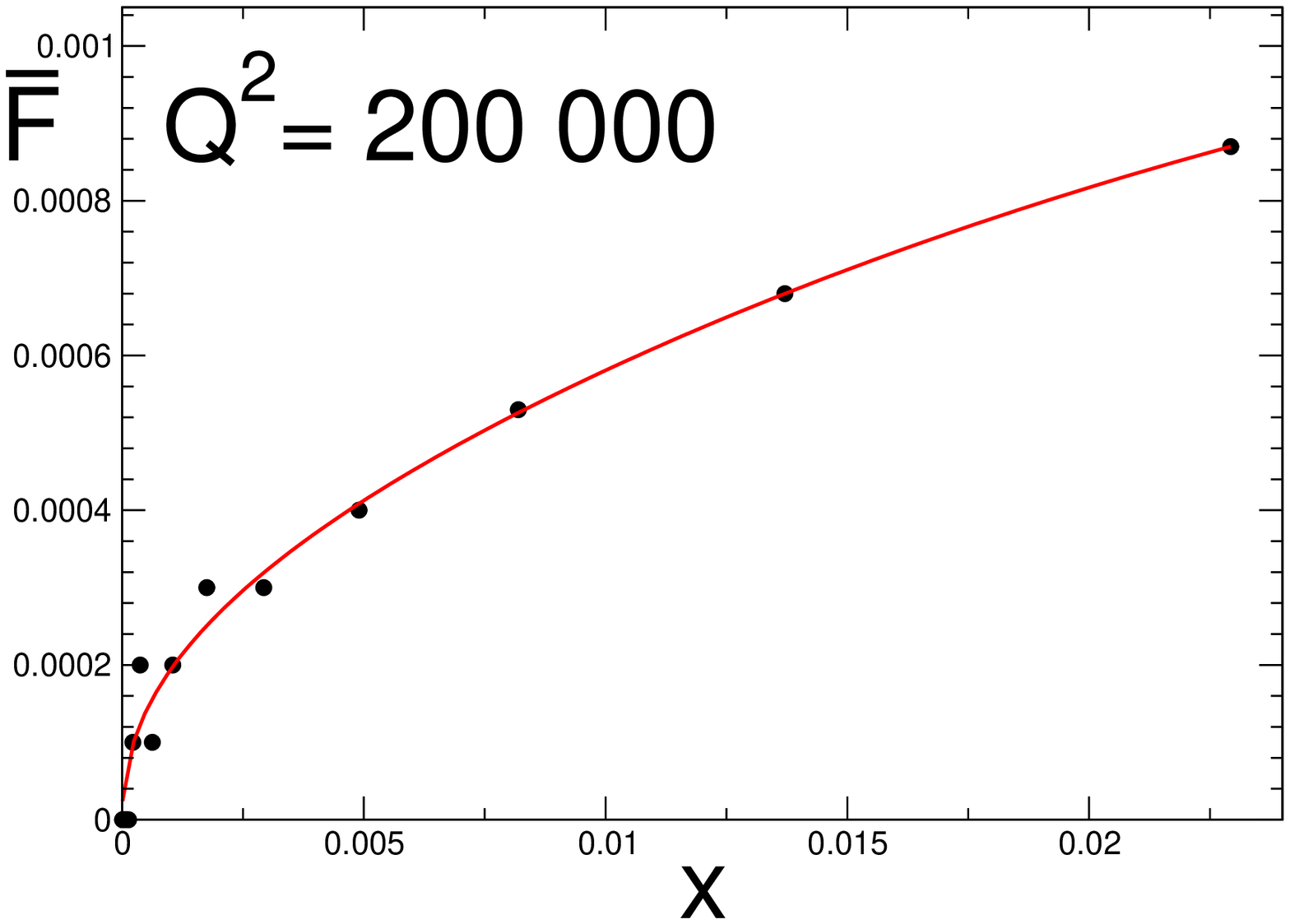}}} 
\caption{Behaviour of the structure function $\bar{F}$ for small $x$. The black dots represent the values extracted from the ABM table while the red curves show the polynomial fits \eqref{eq:FRpoly}. \label{fig:small x}}\end{figure} 

Fig.~\ref{fig:small x} compares the fit (red curves) with the values of $\bar{F}$ obtained from the ABM data (black dots), for various values of $Q^2$, including the lowest and highest ones listed in the table. In the region where the Pomeron dominates, the values of $\bar{F}^p$  and $\bar{F}^n$ are nearly the same. It is difficult to reliably determine the difference between the two from the data on inelastic scattering, even if the DGLAP equations provide a strong theoretical constraint. In the ABM table, the problem also manifests itself directly: for $Q^2>3.5$, the results for $b_\alpha$ exhibit fluctuations which are generated by the limited numerical accuracy of the entries and are visible in Fig.~\ref{fig:small x}. On the other hand, it is questionable, whether the ABM data can be trusted down to $Q^2=1$, because the DGLAP equations rely on perturbation theory. For these reasons we assign an overall relative error of  30\% to the numbers for the difference between the structure functions of proton and neutron obtained from ABM. 
 \begin{figure}
 \resizebox{0.455\textwidth}{!}{%
\includegraphics{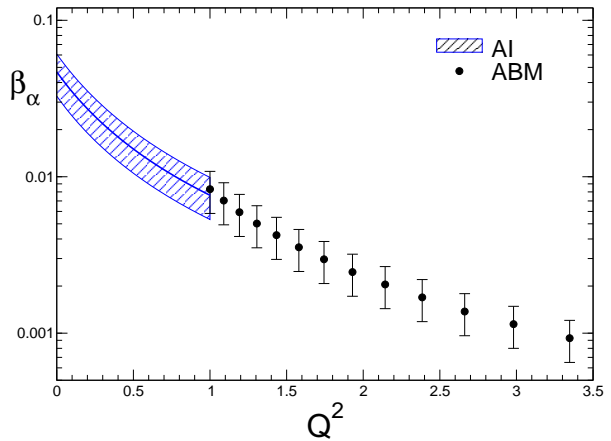}
} 
\caption{Residue of the leading Reggeon. The plot shows the results obtained for the function $\beta_\alpha$, in GeV units. Below $Q^2=1$, the values are based on AI \cite{GVMD}, while above that point, they rely on ABM.  To make them visible despite the very rapid fall-off,  a logarithmic scale is used for $\beta_\alpha$.}
\label{fig:ReggeonResidue}       \end{figure}

Fig.~\ref{fig:ReggeonResidue} compares the Reggeon residue extracted from the ABM table with the values for this quantity obtained from the parametrization of AI. For better visibility, the value of $\beta_\alpha$ is plotted on a logarithmic scale. The figure shows that, at $Q^2=1$, where the two representations meet, the results agree within errors:  the two entirely different sources match, both in sign and in size.  
 \begin{figure}
 \resizebox{0.455\textwidth}{!}{%
\includegraphics*{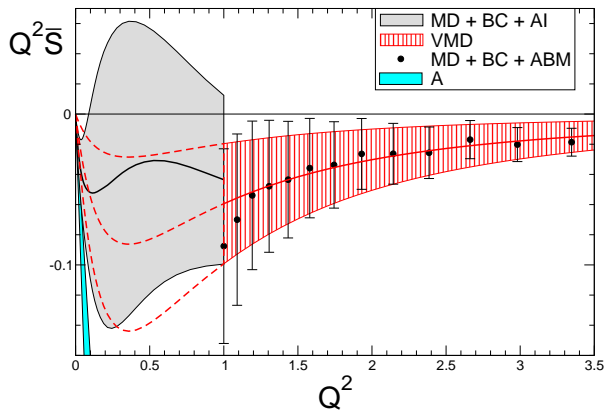}
} 
\caption{Subtraction function at intermediate values of $Q^2$. The black dots represent the values for $Q^2\bar{S}$ obtained from MD, BC and ABM for $Q^2>1$ and the error bars indicate the uncertainty estimates we attach to these. The shaded red band shows the Vector Meson Dominance parametrization of our results in that region, while the dashed red lines represent the extrapolation of this band for $Q^2<1$. The significance of the remaining entries is indicated in the caption of Fig.~\ref{fig:Q2SbarLow}.
 }
\label{fig:Q2SbarIntermediate}       \end{figure} 

Concerning the evaluation of the sum rule for the subtraction function, the only difference compared to the preceding section is that the AI representations for $\bar{F}$ and $b_\alpha$ are replaced by those obtained on the basis of ABM. The black dots in Fig.~\ref{fig:Q2SbarIntermediate} show the outcome -- the error bars are obtained by adding those of the contributions from $W<1.3$ (MD), $1.3<W<3$ (BC) and $W>3$ (ABM) in quadrature. For comparison, the figure also shows the behaviour of the subtraction function for $Q^2<1$, taken over from Fig.~\ref{fig:Q2SbarLow}.

\section{Vector Meson Dominance}
\label{sec:VMD}

As discussed in detail above, the asymptotic freedom of QCD implies that the subtraction function obeys the asymptotic condition \eqref{eq:Sas}:  $\bar{S}\rightarrow C/ Q^{4}$ when $Q^2$ becomes large. The constant $C$ does not represent an unknown, but can be expressed in terms of the mass difference in QCD. Since $C$ is suppressed by chiral symmetry, it is tiny:  $C\approx 6\cdot 10^{-4}\GeV^2$.

In the subtraction function, the numerical noise mentioned above starts becoming visible at $Q^2\approx 3.5$ and, for $Q^2>6$, it hides the signal completely: there, $\bar{S}$ vanishes within errors. 

In order to interpolate between the values of $Q^2$ where the ABM table provides significant information and the region where asymptotics sets in, we make use of the Generalized Vector Dominance Model of Sakurai and Schildknecht \cite{SakuraiSchildknecht}, parametrizing the subtraction function in terms of the contributions from $\rho$, $\omega$ and $\phi$. In the difference between proton and neutron, only the off-diagonal terms survive:
\be\label{eq:SVMD} \bar{S}_{\mbox{\tiny VMD}}(-Q^2)=\frac{1}{m_\rho^2+Q^2}\left\{\frac{c_\omega}{m_\omega^2+Q^2}+\frac{c_\phi}{m_\phi^2+Q^2}\right\}\,.\ee
The asymptotic condition requires the two terms in the bracket to nearly cancel: 
\be\label{eq:cphi}
c_\omega+ c_\phi=C\,.\ee
This leaves a single parameter free, say $c_\omega$. We determine this parameter by fitting the model to the values obtained from MD + BC + ABM in the region $2<Q^2<3.5$.   This range excludes values of $Q^2$ below 2, where the validity of the DGLAP equations is questionable as well as the region $Q^2>3.5$, where the fluctuations show up. The minimum occurs at 
\be\label{eq:comega} c_\omega=-0.74(49)\GeV^2\,.\ee
The red band in Fig.~\ref{fig:Q2SbarIntermediate} shows this fit. 

Since the $Q^2$-dependence of the VMD parametrization reproduces our results very well, the outcome for $m_{\bar{S}}$ is not sensitive to the range used in the fit -- as long as it does not extend into the region $Q^2>6$, where the numerical fluctuations take over. The dashed red lines indicate the behaviour of the VMD parametrization at low values of $Q^2$. Remarkably, although only input for $Q^2>2$ was used, it shows a reasonable behaviour also at low energies. In fact, the central VMD parametrization runs within the error band obtained from the experimental information in the region $Q^2<1$. Evaluating the representation \eqref{eq:SVMD} at $Q^2 = 0$, for instance, and using the relation \eqref{eq:Sbar0alphabeta} between $\bar{S}(0)$ and the polarizabilities, we obtain $\alpha_E^{p - n} - \beta_M^{p - n} = -1.1(7) $. This is about four times more accurate than the available experimental information \eqref{eq:Melendez2} and perfectly consistent with it.

We emphasize, however, that the particular form of the parametrization used to interpolate between low and high values of $Q^2$ does not play a significant role. A parametrization of the form proposed by Erben et al.~\cite{ESTY},
\be\label{eq:SESTY}S_{\mbox{\tiny ESTY}}(-Q^2)=\frac{c_0+C\hspace{0.05em} Q^2  }{(m_0^2+Q^2)^3}\,,\ee
is adequate as well, because it does have the proper asymptotic behaviour. Fixing $m_0$ at the central value used in that reference, treating $c_0$ as a free parameter and fitting it to the values of $\bar{S}$ obtained from MD + BC + ABM in the region $2<Q^2<3.5$, the result for the subtraction function can barely be distinguished from the one obtained with the VMD parametrization.

\section{Asymptotics}
\label{sec:Asymptotics}
 \begin{figure}\begin{center}
\resizebox{0.42\textwidth}{!}{%
\includegraphics{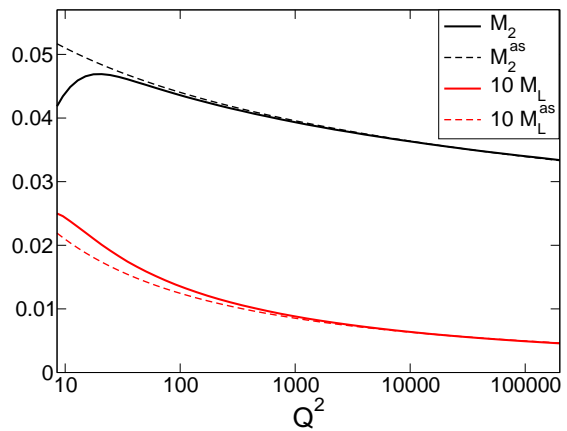}
}
\caption{Moments of the structure functions $F_2$ and $F_L$. The full lines represent the two moments specified in equation \eqref{eq:moments},  while the dashed ones correspond to the asymptotic formulae \eqref{eq:M2}, \eqref{eq:ML} obtained from the operator product expansion. For better visibility, the entries for $M_L$ are stretched with a factor of 10.}
\label{fig:Moments}  \end{center}     \end{figure}
Fig.~\ref{fig:Moments} shows the moments $M_2$ and $M_L$ obtained from the representation of the structure functions we are using -- on a logarithmic scale, so that the entire range covered by the ABM data can be seen. Visibly, the moment $M_L$ is significantly smaller than $M_2$ -- this is to be expected, because the structure function $F_L$ violates Bjorken scaling (at leading order of the perturbative expansion, the structure functions obey the Callan-Gross relation $F_L=0$ \cite{CallanGross}). The dashed lines show the asymptotic behaviour predicted by the operator product expansion. The relations \eqref{eq:M2} and \eqref{eq:ML} fix the momentum dependence of $M_2$ and $M_L$ up to the Wilson coefficients $C_2$ and $C_3$, which represent matrix elements of a spin 2 operator. The results obtained from the ABM analysis are well described by setting $N_f=3$ and using the value $\Lambda_{\QCD}=247\MeV$, for which the leading  order expression for the running coupling constant agrees with observation at $\mu =M_Z$.  Fitting the numerical results for the moments in the range between $Q^2=5\cdot 10^3$ and the upper end of the table provided by ABM, we obtain
\be\label{eq:C2 C3}C_2=0.34 \GeV^2\,,\hspace{2em}C_3=0.072 \GeV^2\,.\ee
Fig.~\ref{fig:Moments} shows that the asymptotic formulae indeed yield a good approximation all the way down to $Q^2\approx 100$. This property is built in: the ABM analysis is based on the DGLAP equations which in turn rely on perturbation theory.  In the region where the effective coupling constant becomes small, the leading terms must dominate. The figure also confirms that $M_L$ disappears more rapidly than $M_2$ by one power of the logarithm, but both moments only fall off very, very slowly.

\begin{figure}
\resizebox{0.42\textwidth}{!}{%
\includegraphics{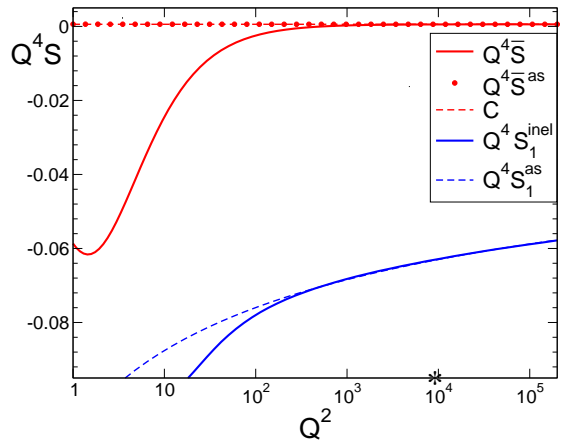}
}
\caption{Asymptotic behaviour of the subtraction function. The red line shows the VMD parametrization of our results for $Q^4 \bar{S}$, while the red dots indicate the asymptotic behaviour that follows from the OPE. The blue lines represent the corresponding
results for the quantity $Q^4S_1^{\inel}$ that plays the same role in traditional analyses of the Cottingham formula ($Q^2$ as well as $\bar{S}$ and $S_1$ are given in GeV units). The star indicates the point where $Q=M_Z$.}
\label{fig:Asymptotics of S}       \end{figure}

Fig.~\ref{fig:Asymptotics of S} shows the behaviour of the structure function $\bar{S}$ at large values of $Q^2$, on a logarithmic scale. The red line represents the VMD parametrization \eqref{eq:SVMD} of our central result. To make the asymptotic behaviour visible, the vertical axis is stretched with the factor $Q^4$. The quantity $Q^4\bar{S}$ approaches the Wilson coefficient $C$, which is determined by the proton matrix elements of the spin 0 operator $\frac{1}{9}(4m_u-m_d)(\ubar u- \dbar d)$ and is indicated by the dashed red line. As discussed in section \ref{sec:Subleading divergence}, $C$ picks up a correction of $O(g^2)$. The red dots represent the values of the function
\be \label{eq:Q4Sbar} Q^4\bar{S}^{\as}=C\left\{1+\frac{g^2}{12 \pi^2}\right\}\,.\ee
The correction is too small to make a visible difference (at the mass of the $Z$-boson, which is marked with a star, it increases the value of $C$ by about 1\%).

Traditionally, the subtraction function is identified with a multiple of $S_1(-Q^2)\equiv T_1(0,-Q^2)$. The relation between this object and the subtraction function we are working with is readily established by comparing the dispersion relations obeyed by $\bar{T}$ and $T_1$. The quantity to compare $\bar{S}$ with is the inelastic part of $S_1$, 
\be\label{eq:S1inel}S_1^{\inel}(-Q^2)\equiv T_1(0,-Q^2)-T_1^{\el}(0,-Q^2)\,.\ee
The comparison of the two dispersion relations yields
\bea\label{eq:DeltaS}S_1^{\inel}(-Q^2)  \al=\al \bar{S}(-Q^2)-\Delta S(-Q^2)\,,\\
\Delta S(-Q^2)\al=\al\frac{m^2}{Q^2}\int_0^{\xth}\hspace{-1em}dx \frac{2F_2(x,Q^2)-F_L(x,Q^2)}{Q^2+m^2x^2}\,.\nonumber\eea
In Fig.~\ref{fig:Asymptotics of S}, our result for $Q^4S_1^{\inel}$ (obtained by subtracting the term $Q^4\Delta S$ from the result for $Q^4 \bar{S}$) is shown as a blue line.
For large values of $Q^2$, the integral over $2F_2-F_L$ becomes proportional to $2M_2(Q^2)-M_L(Q^2)$. With the asymptotic formulae for the moments, the asymptotic behaviour of $S_1^{\inel}$ thus takes the form 
\be \label{eq:Q4S1}  Q^4S_1^{\as}=C\hspace{-0.2em}-\mbox{$\frac{1}{2}$}C_2\left(\!\ln\frac{Q^2}{\Lambda_{\QCD}^2}\!\right)^{\!-d_2}\hspace{-1.5em}
+\mbox{$\frac{1}{4}$}C_3\left(\!\ln\frac{Q^2}{\Lambda_{\QCD}^2}\!\right)^{\!-1-d_2}\hspace{-1.5em}\,. \ee
This shows that, while the asymptotics of $\bar{S}$ is governed by the matrix elements of a scalar operator, $S_1^{\inel}$ picks up additional contributions proportional to the Wilson coefficients $C_2$ and $C_3$, which represent matrix elements of a spin 2 operator. 

The qualitative difference in the asymptotic behaviour of $Q^4\bar{S}$ and $Q^4S_1^{\inel}$ originates in the fact that 
\begin{description}\item{(i)} the approximate chiral symmetry of QCD suppresses the coefficient $C$, while $C_2$, $C_3$ are not suppressed -- they are larger than $C$ by two to three orders of magnitude;
\item{(ii)} while the contribution proportional to $C$ is independent of $Q^2$, those from $C_2$ and $C_3$ fall off logarithmically. 
\end{description} 
Although, eventually, $C$ dominates $Q^4 S_1^{\inel}$ as well, asymptopia is reached only if $Q^2$ is so large that the logarithmic suppression of the spin 2 contributions wins over the chiral suppression of those with spin 0 -- from $Q^2=10^2$ to $Q^2=10^3$, the value of $Q^4 S_1^{\inel}$ only shrinks by about $10\%$. 

The counter term $\Delta m^\Lambda$ only removes the leading and subleading divergences associated with $C$. The additional divergence proportional to $C_2$ does not have anything to do with renormalization and is of purely technical nature: equation \eqref{eq:M2} shows that the same divergence also shows up in the asymptotic behaviour of $T_2$.  In the sum of the contributions from $S_1^{\inel}$ and $T_2$, the spin 2 divergences cancel \cite{Collins}. It is difficult, however, to specify the contribution from $S_1^{\inel}$ by itself: the asymptotic formula \eqref{eq:Q4S1} shows that this contribution diverges unless the non-leading term proportional to $C_2$ is removed as well as the leading one. Our framework avoids these problems.
 
\section{Numerical evaluation of the mass difference}
\label{sec:Numerics}
\subsection{\texorpdfstring{Form factors, $m_{\el}$}{Form factors, m\string_(\el)}}
\label{sec:Form factors}

The elastic contribution to the e.m.~part of the mass difference is determined by the form factors. In early work, the experimental information about these was adequately described by the dipole formulae (see e.g.~appendix A of \cite{GHLR}).
They yield  $0.63\MeV$ for the proton and $-0.13\MeV$ for the neutron, so that the elastic contribution to the self-energy difference amounts to  $m_{\el}=0.76\MeV$ \cite{GL1975}. In the meantime, the precision to which the form factors are known has increased significantly \cite{Formfactors,Kelly,Ye2018,Borah}. Using this information, we obtain 
\be\label{eq:mel}m_{\el}=0.75\pm 0.02\MeV\,.\ee
The error bar covers the results obtained with the three parametrizations of \cite{Formfactors,Kelly,Ye2018}. This 
indicates that, in the difference between the e.m.~self-energies of proton and neutron, the departures from the dipole formulae only generate a change of the order of a percent. The uncertainties in the result for the mass difference generated by the elastic part are totally neglible compared to those from the inelastic contributions.

\subsection{Contribution from the subtraction function}
\label{sec:Contribution from the subtraction function}
The contribution to $m_{\bar{S}}$ depends on the scale $\mu$ used in the e.m.~renormalization of the quark masses. For definiteness, we use $\mu=\mu_2\equiv 2\GeV$. If $\mu$ is taken differently, the mass difference changes by $2NC\ln(\mu/\mu_2)$. 

In the region $0<Q^2<1$ our representation of the subtraction function is based on the parametrizations MD, BC and AI (gray band in Fig~\ref{fig:Q2SbarIntermediate}). Inserting this representation in formula \eqref{eq:mSbar} we obtain
\be\label{eq:mSbarLow}m_{\bar{S}}\mbox{\small${(0<\!Q^2\!<\!1)}$}=-0.034(68)\MeV\,.\ee
The central value is negative and reduces the elastic contribution by about $5\%$. The error is twice as large, however, so that small positive contributions from this region are not excluded. 

Since the integrand of $m_{\bar{S}}$ is proportional to $Q^2\bar{S}$, small values of $Q^2$ are suppressed; the fictitious spike occurring there in the parametrization of BC (see Figs.~3--5 in \cite{GHLR}) does not affect the result very strongly, but an improved analysis of the structure functions in the resonance region above the $\Delta(1232)$ would allow reducing the quoted uncertainty. 
 
In the region $1<Q^2<\infty$, we use the VMD parametrization of $\bar{S}$ and obtain  
\bea\label{eq:mSbarHigh}\al\al m_{\bar{S}}\mbox{\small $(1\!<\!Q^2\!<\!2)$}=-0.040(27)\MeV\,,\\
\al\al m_{\bar{S}}\mbox{\small $(2\!<\!Q^2\!<\!\infty)$}= -0.092(61)\MeV\,.\nonumber\eea
To account for the correlations between the contributions from the various regions, we determine the net error in $m_{\bar{S}}$ by evaluating the integral in equation \eqref{eq:mS} for the upper and lower edges of the error band. This leads to
\be\label{eq:mSbartot}m_{\bar{S}}=-0.17(16)\MeV\,.\ee

\subsection{Contributions from the dispersion integrals}
Finally, we evaluate the convergent integrals $m_{\bar{F}}$, $m_{F_2}$  in equations \eqref{eq:mFbar} and \eqref{eq:mF2}. In these integrals, the small $x$ region does not require special care. As mentioned above, the angular integration suppresses the contributions from the  deep inelastic region. In fact, a very strong suppression also occurs at low values of $Q^2$. Numerically, these integrals are tiny: 
\bea\label{eq:mF}
m_{\bar{F}}\al=\al -0.0004(4) \MeV\,,\\
m_{F_2}\al=\al -0.0039(10)\MeV\,.\nonumber\eea

\subsection{\texorpdfstring{Result for $m_{\QED}$ and $m_{\QCD}$}{Result for m \string_ (QED) or m_string_(QCD)}}
Collecting the various contributions, the part of the proton-neutron mass difference that is due to the e.m.~interaction becomes \be\label{eq:meQED} m_{\QED}= 0.58\pm 0.16\MeV\,.\ee
The observed mass difference then yields
\be\label{eq:mQCD}m_{\QCD}=-1.87\mp 0.16\MeV\,.\ee 
The result for $m_{\QCD}$ provides a more precise estimate for the leading Wilson coefficient: 
\be\label{eq:Cexp}C= 5.7(1.1)\cdot 10^{-4}\GeV^2\,.\ee
We have repeated the entire calculation with this input instead of the crude estimate used for this constant. At the quoted accuracy, the results stay put.  
\section{Comparison with Lattice calculations}
\label{sec:Lattice}
 
\begin{figure}
  \begin{center}
\resizebox{0.49\textwidth}{!}{%
    \includegraphics{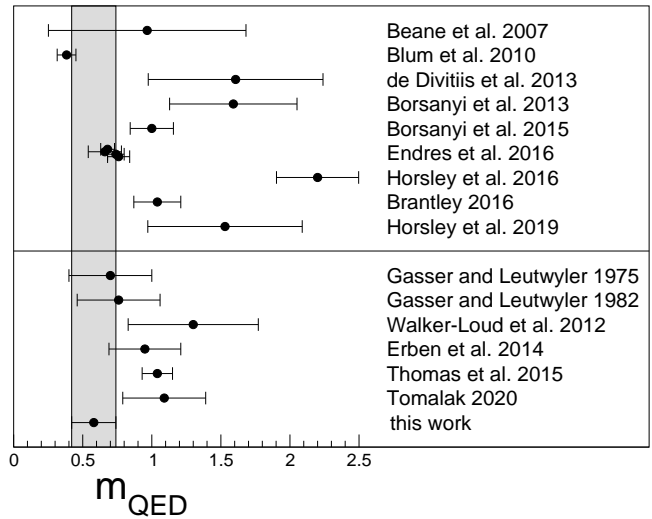}
}
\end{center}
\caption{Electromagnetic part of the mass difference between proton and neutron (MeV units). The upper part shows lattice results  \cite{Beane:2006fk,Blum:2010ym,deDivitiis:2013xla,Borsanyi:2013lga,Borsanyi2015,Endres:2015gda,Horsley:2015vla,Brantley:2016our,Horsley:2019wha}, the lower part contains results obtained with the Cottingham formula \cite{GL1975,GL1982,WCM,ESTY,TWY,Tomalak}, including the outcome of our analysis.}
\label{fig:mQED}\end{figure}

Within QCD, the lattice approach allows a determination of the mass spectrum with steadily increasing precision, not only for the mesons but also for the more difficult case of the baryons. The inclusion of the e.m.~interaction gives rise to a serious problem, however, because this interaction is of long range -- enclosing the system in a box distorts the results through finite size effects that need to carefully be sorted out. In comparison  with the extensive documentation available for lattice determinations of the quark masses within QCD, the literature containing numerical results for $m_{\QED}$ is rather scarce. Fig.~\ref{fig:mQED} collects the results we found. Visibly, the likelihood for the results listed to represent statistically independent measurements of the same physical quantity is quite small. Indeed, not all of the errors shown include an estimate for the systematic uncertainties. Also, not all of the listed papers have appeared in print. Some of the results are obtained from a calculation that simulates QCD+QED, others stay within QCD, calculate the part due to the difference between $m_u$ and $m_d$ and determine the part that comes from the e.m.~interaction by comparing the calculated part with the experimental value.  It is well-known that the splitting into two parts depends on the convention used, but this is a theoretical problem that does not require numerical simulations. 

Our numerical result for $m_{\QED}$ is dominated by the elastic contribution; the remainder is significantly smaller and negative. The most recent lattice results listed in Fig.~\ref{fig:mQED} are instead larger than the elastic contribution: the remainder is positive and comparable to the elastic term. Clearly, our result is not consistent with that. 
\section{Comparison with other evaluations of the Cottingham formula}
\label{sec:Comparison}

We are aware of four recent estimates for the proton-neutron mass difference based on evaluations of the Cottingham formula:  Walker-Loud, Carlson and Miller (WCM) \cite{WCM,Walker-Loud2018}, Erben, Shanahan, Thomas and Young (ESTY) \cite{ESTY}, Thomas, Wang and Young (TWY) \cite{TWY}  and Tomalak  \cite{Tomalak}. The first three propose models for the subtraction function $S_1(-Q^2)$, using the experimental information concerning the difference between the magnetic polarizabilities of proton and neutron to determine the value of $S_1(0)$ and making a simple algebraic ansatz for the momentum dependence. A detailed comparison of the models proposed by WCM and ESTY with the results obtained from Reggeon dominance at low values of $Q^2$ can be found in \cite{GHLR}.  

Tomalak  \cite{Tomalak}  also uses the available experimental information about the magnetic polarizabilities, but instead of making an ansatz for the momentum dependence of the subtraction function, he calculates it on the basis of the assumption that -- once the contributions from the Reggeons are removed -- the amplitude $\hat{T}_1=q^2T_1+\nu^2T_2$ obeys an unsubtracted dispersion relation \cite{TV}. Although this assumption resembles Reggeon dominance, we consider it very unlikely that it is correct.  For $q^2 = 0$, for instance, the amplitude $\hat{T}_1$ reduces to $\nu^2T_2$. The asymptotic behaviour of this quantity was investigated by Damashek and Gilman \cite{DG} and, independently, by Dominguez, Ferro Fontan and Suaya \cite{Dominguez}. Their work indicates that $f=\nu^2 (T_2-T_2^{\indR})$  tends to a nonzero constant when $\nu$ becomes large. The assumption used in \cite{Tomalak} instead implies that $f$ tends to zero. At any rate, this hypothesis implies a constraint on the imaginary part of $T_2$ at $q^2=0$, i.e.~on the cross section of photoproduction: it leads to a sum rule that requires an integral over the cross section to cancel the Thomson term.  We do not know of an argument that would support this assumption.

Incidentally, the assumption used in \cite{Tomalak} corresponds to a special case of the universality hypothesis of Brodsky, Llanes-Estrada and Szczepaniak \cite{BLS,BCG,BCG1}, who do not impose the condition that the difference $\hat{T}_1-\hat{T}_1^R$ tends to zero for $\nu\rightarrow \infty$, but postulate that it becomes independent of $q^2$. We cannot see any reason for this to be the case in QCD (see also \cite{Creutz,Mueller2015,Mueller2016}). 

\subsection{Contributions from the elastic part}
\label{sec:Elastic}
Since $T_2$ obeys an unsubtracted dispersion relation, the corresponding Born term is readily obtained by saturating the dispersion integral with the contributions from the nucleon poles. For $T_1$, however, the Born term is not unique -- various different expressions are used in the literature. They all obey a subtracted dispersion relation, but differ in the choice of the subtraction function. 

Dispersion theory offers a unique solution: since analytic functions are determined by their singularities and their behaviour at infinity, it suffices to impose the condition that the Born term vanishes for $\nu\rightarrow\infty$.  We refer to the resulting expression as the elastic part of the amplitude. It is explicitly given in formula \eqref{eq:Born term} (the unsubtracted dispersion relation used to specify the Born term for $T_2$   automatically ensures that it disappears if $\nu$ becomes large).  Accordingly, the elastic part of $m_{\QED}$, which we denote by $m_{\el}$,  is an unambiguous notion as well. It is obtained by replacing the amplitudes in \eqref{eq:mgammaLambda} by their elastic parts and removing the cutoff -- the elastic contributions are convergent.  

WCM \cite{WCM} instead represent the elastic part of the mass difference with two terms.\footnote{For a critical examination of their line of reasoning, we refer to Appendix E in \cite{GHLR} and to \cite{HLCD15,Hoferichter:2019jhr}.} The sum of the two, $\delta M_{\el}+\delta M_{\el}^{\mbox{\tiny sub}}$, differs from $m_{\el}$ by
\be\label{eq:MelWCM}\Delta m_{\el}  =-\frac{3\alphaem m}{2\pi}\hspace{-0.3em}\int_0^\infty\hspace{-0.7em}dQ^2Q^2\,\frac{(G_E-G_M)^2}{(4m^2+Q^2)^2} \,.\ee
Numerically, $\Delta m_{\el}$ is small: using the parametrization of Kelly \cite{Kelly}, we obtain $\Delta m_{\el}^p=-0.051\MeV $, $\Delta m_{\el}^n=-0.064\MeV$. In the difference between proton and neutron, these numbers even partly cancel. 

At the precision at which the nucleon form factors can nowadays be measured, it matters  whether the standard expression for $m_{\el}$ or the quantity $m_{\el}+\Delta m_{\el}$ is determined. For the decomposition \eqref{eq:decomposition} to be valid, it is essential that the nucleon form factors exclusively occur in $m_{\el}$ -- any other representation of the elastic part must be compensated by a corresponding correction in the term arising from the subtraction function. 

\subsection{Contributions from the subtraction function}
\label{sec:Contributions}
As demonstrated in the preceding sections, the inelastic contributions to $m_{\QED}$ are totally dominated by the one from the subtraction function $\bar{S}$. The differences in the values quoted for the elastic contributions are small compared to those from inelastic processes. Hence we can compare the various determinations of the mass difference that rely on dispersion theory by comparing the corresponding representations for $\bar{S}$. 

\begin{figure}
\resizebox{0.48\textwidth}{!}{%
 \includegraphics{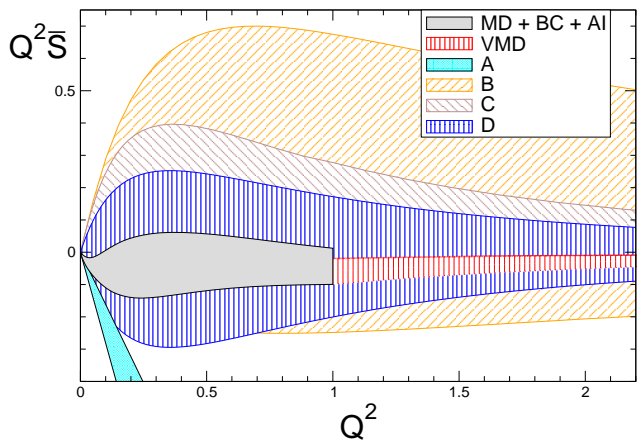}
}
\caption{Subtraction function versus $Q^2$, in GeV units. The bands labeled MD+BC+AI and VMD indicate our representation in the region below and above $Q^2=1$, respectively. The wedge A depicts the tangent at $Q^2=0$ obtained with the magnetic polarizability of \cite{Bignell},
 B: model \cite{WCM},
 C: model \cite{ESTY},
 D: parametrization of \cite{ESTY} applied to $\bar{S}$.
 }
\label{fig:Comparison}       \end{figure}

The bands labeled B and C in Fig.~\ref{fig:Comparison} show the models for the subtraction function of WCM \cite{WCM} and ESTY \cite{ESTY}, respectively. They are obtained from the representations proposed for $S_1$ in these references, merely converting numbers for $S_1^{\inel}$ into numbers for $\bar{S}$ by means of equation \eqref{eq:DeltaS}. The width of the bands exclusively shows the uncertainties arising from the experimental information used for the magnetic polarizabilities -- those associated with the freedom in the choice of the model would widen it further. In the $Q^2$ range shown in the figure, both models are consistent with our analysis, but come with significantly larger errors (as the lower edge of band C runs within our band of uncertainties, it cannot be seen in Fig.~\ref{fig:Comparison}).

The input used in models B and C for the value of the subtraction function at $Q^2=0$  is the same -- it is based on the experimental determination of the polarizabilities of the nucleon. At small values of $Q^2$, our uncertainties are smaller because the predictions obtained from Reggeon dominance for the polarizabilities of the neutron \cite{GHLR} are more precise than the experimental values. An improved measurement of the polarizabilities would be most welcome as it would subject Reggeon dominance to an important test. In this connection, we also refer to the new lattice results on the magnetic pola\-rizabilities discussed in section \ref{sec:Polarizabilities}.

At large values of $Q^2$, the uncertainty band attached to model C is more narrow than the one of B, because the parametrization is improved: asymptotically, model C does reproduce the leading term in the operator product expansion of $S_1^{\inel}$. As can be seen in Fig.~\ref{fig:Asymptotics of S}, however, the nonleading spin 2 contributions disappear only extremely slowly. In the parametrization of model C, these are neglected. 

The net result obtained with model C for $\beta_M^{p-n}=-0.5(1.6)$ is $m_{\QED}^C=0.95(25)\MeV$ \cite{ESTY}. The corresponding outcome for the contribution from the subtraction function is obtained by removing the elastic part as well as those from the convergent dispersion integrals. With the entries for the elastic contributions listed in Table I of \cite{ESTY} and the values given in equation \eqref{eq:mF} for the tiny terms $m_{F_2}$ and $m_{\bar{F}}$, this yields 
\be\label{eq:mSbarC}
m_{\bar{S}}^C=0.19(25)\,.\ee 
The value obtained by integrating the subtraction function of model C only over the low energy region is nearly the same:  
$m_{\bar{S}}^C(Q^2\!<\!2)=0.20(29)\MeV$. This indicates that in the evaluation of model C in \cite{ESTY}, the contribution from 
$Q^2>2$ is nearly cancelled by the counter term,
but we cannot verify this within our own framework. Since the parametrization of $S_1$ used in model C neglects the non-leading contributions in the asymptotic formula \eqref{eq:Q4S1}, it does not make sense to insert the corresponding representation for $\bar{S}$ in the expression \eqref{eq:mSbar} for $m_{\bar{S}}$ -- the integral diverges. Also, the blue line in Fig.~\ref{fig:Asymptotics of S} shows that for $S_1^{\inel}$, asymptopia sets in extremely slowly, because the contributions generated by the short distance singularities of spin 2 fall off only logarithmically.  

The numerical results for the subtraction function used by TWY and Tomalak are very similar to model C and they also lead to similar results for the e.m.~part of the mass difference: $m_{\QED}=1.04(11)\MeV$ (TWY) and  $m_{\QED}=1.09(30)\MeV$ (Tomalak). The difference mainly arises from the input used for $\beta_M^{p-n}$. Note that the value $\beta_M^{p-n}=-1.12(40)$ used by TWY comes with a remarkably small error and disagrees with the Reggeon dominance prediction \eqref{eq:betaGHLRM} by 2.5 $\sigma$. This is puzzling, because the determination of $\beta_M^{p-n}$ in TWY is based on the lattice data of Blum et al.~\cite{Blum:2010ym} -- as shown in Fig.~\ref{fig:mQED}, these data are perfectly consistent with the range for $m_{\QED}$ obtained from Reggeon dominance.

The ambiguities related to the fact that the function $S_1^{\inel}$ approaches asymptotics only very slowly do not arise if the parametrization of model C is used to represent $\bar{S}$ rather than $S_1^{\inel}$. We refer to this option as model D: the momentum dependence of $\bar{S}$ is described by the function specified in equation \eqref{eq:SESTY}, $m_0$ is identified with the scale $m_0^2=0.71\GeV^2$ occurring in the dipole representation of the nucleon form factors \cite{Walker-Loud2018}  and the parameter $c_0$ is fixed with the experimental value $\bar{S}(0)=-0.2(2.6)\GeV^{-2}$ given in equation \eqref{eq:Sbar0M}, which is based on the determination of the polarizabilities in \cite{Melendez}.  The blue shaded region in Fig.~\ref{fig:Comparison} shows that the subtraction function obtained with this variant of the models proposed in \cite{WCM,ESTY,TWY} agrees perfectly well with our analysis, but comes with a much larger error. Inserting  the parametrization of model D in formula \eqref{eq:mSbar}, we obtain  
\be\label{eq:mSbarD}m_{\bar{S}}^D=-0.05(61)\MeV\,. \ee
The region $Q^2>2$ does not contribute much to the central value, but is responsible for a substantial fraction of the error:  
$m_{\bar{S}}^D(Q^2\!>\!2)=-0.02(28)\MeV$.  

C and D have the same behaviour at very small and very large values of $Q^2$ -- they only differ in the form of the interpolation used in between. The example shows that -- if only the leading terms in the OPE of the subtraction function are accounted for, the outcome is very sensitive to the form of the interpolation: replacing C by D lowers the central value of $m_{\bar{S}}$ by $0.24 \MeV$ and thus lowers the outcome for the central value of the mass difference to $m_{\QED}=0.71\MeV$. This is within the uncertainty range attached to our result \eqref{eq:meQED}. The sensitivity to the form chosen for the interpolation arises because the subtraction function $S_1$ reaches asymptotics only very slowly.

Our analysis is not affected by this ambiguity, because we calculate the subtraction function in the region $Q^2<3.5$ on the basis of the experimental information about the structure function $\bar{F}$ and rely on the theoretical information about the asymptotics only at higher energies. As pointed out in section \ref{sec:VMD}, the contribution arising from the region $Q^2>2$ is nearly independent of the form of the parametrization used there, provided only that it obeys the theoretical constraints imposed by asymptotic freedom.  
\section{Summary and conclusions}
\label{sec:Summary}

\hspace{1em}1.~Dispersion theory determines the amplitude in terms of its physical singularities (poles, cuts), provided the asymptotic behaviour is known. The use of amplitudes that contain kinematic zeros is best avoided, because these make it very difficult to sort out the asymptotic behaviour. We work with the invariant amplitudes introduced by Cottingham which do not contain such deficiencies and which we denote by $T_1,T_2$.

2.~In the framework of dispersion theory, the elastic part of $T_1,T_2$ is an unambiguous notion, determined by the requirement that it is analytic except for the poles generated by the elastic reaction and disappears when $\nu \rightarrow\infty$. Accordingly, the elastic contribution to the Cottingham formula is unambiguous. 

3.~As we do not know the error matrix occurring in the determinations of the form factors, we are not in a position to give a reliable estimate for the uncertainties in $m_{\el}$. We instead rely on the results obtained with the three different parametrizations in \cite{Formfactors,Kelly,Ye2018}, which are covered by
\be \label{eq:mel3}m_{\el}=0.75\pm0.02\MeV\,.\ee
A determination of $m_{\el}^p$, $m_{\el}^n$ and $m_{\el}^{p-n}$ on the basis of the information about the nucleon form factors available today would reduce the error considerably, but at the precision to which the inelastic contributions can currently be determined, the uncertainty quoted in \eqref{eq:mel3} is too small to affect the error estimate attached to our result for $m_{\QED}$.

4.~The leading terms of the operator product expansion of the Compton amplitude involve contributions arising from short distance singularities related to operators of spin 0 as well as spin 2. We make use of the fact that the leading spin 2 contributions to $T_1$ and $T_2$ only differ in normalization: in the combination $\bar{T}\equiv T_1+\frac{1}{2}T_2$, they drop out. Replacing the pair $\{T_1,T_2\}$ by   $\{\bar{T},T_2\}$ simplifies the analysis considerably.

5.~A further simplification occurs if the dispersion relation for $\bar{T}$ is not subtracted at $\nu=0$, but at $\nu=\frac{1}{2}\sqrt{q^2}$.
This ensures that the contributions from the dispersion integrals over $\Im \bar{T}$ and $\Im T_2$ both contain the factor $\nu^2-\frac{1}{4}q^2$. The point here is that in the Cottingham formula, only the angular average matters. Since the angular average of $\nu^2-\frac{1}{4}q^2$ vanishes, the contributions from the dispersion integrals are suppressed -- numerically, these contributions are tiny. In our decomposition of the amplitude, only the elastic term and the integral over the subtraction function can generate significant contributions to the mass difference.  

6.~The quarks and gluons reggeize. The exchange of Reggeons generates moving poles. For large values of $\nu$ at fixed $q^2$, a Reggeon contributes with $\bar{T}\propto \nu^\alpha$ and $T_2\propto \nu^{\alpha-2}$, where $\alpha$ is the value of the trajectory $\alpha(t)$ at $t=0$. Since there are trajectories with $\alpha >0$, the dispersion relation for $\bar{T}$ must be subtracted. The one for $T_2$ does not require a subtraction. 

7.~We assume that the asymptotic behaviour of $\bar{T}$ is dominated by the contributions from the Reggeons, which we denote by $\bar{T}^{\indR}$. More precisely, we require that $\bar{T} -\bar{T}^{\indR}$ tends to zero when $\nu\rightarrow\infty$ and refer to this assumption as {\it Reggeon dominance}. A nonzero limiting value would represent a fixed pole -- we are thus assuming that  Reggeization is complete and only moving poles occur. Note that the dispersion relations for $\bar{T}$ and $T_2$ imply the presence of contributions that fall off with the power $\nu^{-2}$. In $T_2$, these contributions correspond to a fixed pole at $\alpha=0$ -- Reggeon dominance is perfectly consistent with fixed poles of this sort.

8.~Reggeon dominance implies a sum rule that determines the subtraction function $\bar{S}$ in terms of the structure function $\bar{F}$. The explicit expression given in \eqref{eq:Sum rule} shows that neither the nucleon form factors nor the structure function $F_2$ enter. A variant of this sum rule was proposed by Elitzur and Harari, long ago \cite{ElitzurHarari}, on the basis of duality and finite energy sum rules.

9.~The value of $\bar{S}(q^2)$ at $q^2=0$ is related to the polarizabilities of the nucleon. As is well known, the sum of the electric and magnetic polarizabilities is determined by a sum rule involving the cross section for photoproduction. Reggeon dominance implies separate sum rules for the electric and magnetic polarizabilities. The prediction obtained for the difference between the magnetic polarizability of proton and neutron \cite{GHLR} is in agreement with experiment, but this represents only a rather weak test of Reggeon dominance, because the uncertainties in the experimental result are rather large. The errors attached to the recent lattice result of \cite{Bignell} are much smaller -- it is encouraging that Reggeon dominance passes this more stringent test as well. More work on the polarizabilities, particularly those of the neutron, would be most welcome.  

10.~Theory fixes the asymptotic behaviour of the subtraction function: if $Q^2$ becomes large, $\bar{S}$ tends to $C/Q^4$, where the constant $C$ is given by the proton matrix element of the operator $\frac{1}{9}(4m_u-m_d)(\bar{u}u-\bar{d}d)$. This also holds for  $S_1(q^2)=T_1(0,q^2)$, the subtraction function commonly used in dispersive analyses of the Compton amplitude, but  the short distance singularities related to operators of spin 2 generate a significant difference in the asymptotic behaviour. Fig.~\ref{fig:Asymptotics of S} compares the momentum dependence of $S_1$ and $\bar{S}$ on a logarithmic scale and shows that, in contrast to $\bar{S}$, the asymptotics of $S_1$ sets in only very, very slowly. 

11.~An important part of the calculation concerns the determination of the residue of the Reggeon with the quantum numbers of the $a_2$, which dominates the asymptotic behaviour of the difference between the amplitudes of proton and neutron. Fig.~\ref{fig:ReggeonResidue} shows that the result obtained at low values of $Q^2$ from the Regge representation of \cite{GVMD} matches the outcome of the Regge fit to the numerical ABM table remarkably well.

12.~With the values for the subtraction function obtained from the solution of the sum rule, our net result for the e.m.~part of the mass difference between proton and neutron reads
\be\label{eq:mQED}m_{\QED}^{p-n}= 0.58\pm 0.16\MeV\,.\ee
The conclusions reached in Ref.~\cite{GL1975} are thus confirmed: $m_{\QED}$ is dominated by the elastic contribution. The uncertainty in the result obtained forty five years ago, $m_{\QED}= 0.7(3)\MeV$ \cite{GL1975}, is reduced by about a factor of two. In the present analysis, the uncertainty is predominantly due to the contributions from the resonance region above the $\Delta(1232)$. It could be reduced by an improved experimental determination of the structure functions in that region, particularly for the neutron.

13.~It is by no means puzzling that the inelastic contributions are so small: (a) the angular integration suppresses the contributions from the dispersion integrals, (b) at large values of $Q^2$, the subtraction function is nearly the same for proton and neutron -- in the chiral limit, there is no difference, (c) in the region where Reggeon exchange dominates, the leading term, the Pomeron, is the same, (d) isospin symmetry ensures that the most important resonance, the $\Delta(1232)$, contributes equally to proton and neutron and (e) the leading terms of the chiral perturbation series are also the same.

With the experimental value of the mass difference, the above result implies that the part due to the difference between $m_u$ and $m_d$ is given by
\be m_{\QCD}^{p-n}=-1.87\mp 0.16\MeV\,.\ee

14.~The lattice results for these quanitities did not yet reach a level of coherence to be covered by the FLAG report, but the method is steadily being improved and, in the long run, should provide reliable numbers. Fig.~\ref{fig:mQED} indicates that the most recent lattice values
 are larger than the outcome of the present work.  If the value of $m_{\QED}^{p-n}$ should turn out to be larger than $1\MeV$, we would have to conclude that the Compton amplitude does not fully reggeize: the amplitude $T_1$ would then contain a fixed pole that invalidates the Reggeon dominance hypothesis. We would then be left with a puzzle: what is the physical origin of this fixed pole?  

16.~The evaluations of the Cottingham formula in \cite{WCM,ESTY,TWY} lead to values for $m_{\QED}$ around $1\MeV$. In these references, a simple algebraic ansatz is used to parametrize $S_1^{\inel}$, the inelastic part of the subtraction function $T_1(0,q^2)$. Fig.~\ref{fig:Asymptotics of S} shows that, in contrast to these parametrizations, $S_1^{\inel}$ approaches asymptotics only extremely slowly. 

The mismatch with the asymptotics disappears if the ansatz is assumed to be valid for $\bar{S}$ rather than $S_1^{\inel}$. The central value obtained for $m_{\QED}$ then drops by $0.24\MeV$ and winds up slightly below the elastic contribution, in agreement with what we find. On the other hand, quite apart from the sensitivity to the precise form of the assumptions underlying those models, the uncertainties in the result for $m_{\QED}$ are much larger than ours, because the experimental determination of $\beta_M^{p-n}$, which plays a key role in that approach, is subject to large uncertainties.

\begin{acknowledgements}
We thank Johannes Bl\"umlein for providing us with numerical tables for the structure functions based on the ABM solutions of the DGLAP equations, Thomas Becher for information about the anomalous dimensions in QCD+QED, Vadim Lensky and Vladimir Pascalutsa for a Mathematica notebook concerning the representation of the Compton amplitude in $\chi$PT and Ryan Bignell, Irinel Caprini, Stefano Carrazza, Gilberto Colangelo, Cesareo Dominguez, Franziska Hagelstein, Bastian Kubis, Ulf-G.~Mei{\ss}ner, Sven-Olaf Moch, Gerrit Schierholz and Ignazio Scimemi for comments and useful information. 
A.R. acknowledges the support from the DFG (CRC 110 ``Symmetries 
and the Emergence of Structure in QCD''), as well as from
Volkswagenstiftung under contract no. 93562. 

\end{acknowledgements}

\noindent Authors' comment: The grace files used for the figures can be found as ancillary files 
in the archive version of the present article under the link https://arxiv.org/abs/2008.05806.

\appendix
\section{Form of the Wilson coefficient for spin 2}
\label{sec:Wilson2}
Lorentz invariance implies that a tensor $C_{\mu\nu\alpha\beta}(q)$ that is symmetric under $\mu\leftrightarrow\nu$ and $\alpha\leftrightarrow\beta$ and only depends on the four-vector $q^\mu$ is of the form: 
\bea \al\al C_{\mu\nu\alpha\beta}(q)=a\, g_{\mu\nu}g_{\alpha\beta}+b\,\{g_{\mu\alpha}g_{\nu\beta}+g_{\nu\alpha}g_{\mu\beta}\}\\
\al\al+c\, g_{\mu\nu}q_\alpha q_\beta+ d\, g_{\alpha\beta}q_\mu q_\nu +e\,\{g_{\mu \alpha}q_\nu q_\beta+g_{\mu\beta}q_\nu q_\alpha\no
\al\al+g_{\nu\alpha}q_\mu q_\beta+ g_{\nu\beta}q_\mu q_\alpha \}+f\,q_\mu q_\nu q_\alpha q_\beta\,,\nonumber\eea
where $a,b,c,d,e,f$ can only depend on $q^2$. Current conservation, $q^\mu C_{\mu\nu\alpha\beta}(q)=0$, fixes $a,b,c$ in terms of $d,e,f$:
\be\label{eq:abc} a+q^2 d =0\,,\hspace{1em}b+q^2e = 0\,,\hspace{1em}c+2e+q^2f = 0\,.\ee
On account of the tracelessness of $O^{\alpha\beta}$, the coefficient $d$ drops out in the sum
$C_{\mu\nu\alpha\beta}(q)O^{\alpha\beta}$. To simplify the notation, we replace the coefficients $e,f$ by $c_2,c_3$, with $e=\frac{1}{2} c_2$, $f=c_3$.  The contribution from an operator of spin 2 to the OPE then takes the form:\footnote{Hill and Paz \cite{HillandPaz} use a different normalization. In our notation, they work with $T_{\mu\nu}^{\text{\tiny HP}}= 2T_{\mu\nu}$ and normalize the spin 2 operator formed with the derivatives of the quark fields differently: $O_{\alpha\beta}^{f_2\,\text{\tiny HP}}= \frac{1}{2}O_{\alpha\beta}^{f_2}$. The Wilson coefficients are related  by  $c_1^f=c_1^{\text{\tiny{HP}}}/(2q^4)$, $c_2^f=-c_3^{\text{\tiny{HP}}}/(4q^4)$, $c_3^f=(c_2^{\text{\tiny{HP}}}+c_3^{\text{\tiny{HP}}})/(4q^6)$.} 
\bea\label{eq:Cc23}\al\al C_{\mu\nu\alpha\beta}(q)O^{\alpha\beta}=c_3\,(q_\mu q_\nu-g_{\mu\nu}q^2) O_{\alpha\beta}q^\alpha q^\beta\\
\al\al\hspace{1em} +c_2\, (g_{\mu\alpha}O_{\nu\beta}+g_{\nu\alpha} O_{\mu\beta}-g_{\mu\nu}O_{\alpha\beta}) (q^\alpha q^\beta-\mbox{$\frac{1}{2}$}g^{\alpha\beta}q^2)\,.\nonumber \eea
This shows that, while Lorentz invariance and current conservation fix the Wilson coefficients belonging to operators of spin 0 in terms of a single function $c_1(q^2)$, those associated with operators of spin 2 involve two such functions: $c_2(q^2)$ and $c_3(q^2)$. 
 
\section{Operator product expansion for free quarks}
\label{sec:OPE for free quarks}
\setcounter{equation}{0}
For free quarks, the time-ordered product of two currents can be decomposed as 
\bea \label{eq:TN}T j_\mu(x)j_\nu(y)\al=\al\sumf Q_f^2\,\mbox{tr}\{\gamma_\mu S^f(x-y)\gamma_\nu S^f(y-x)\}{\bf1}\no
\al+\al\frac{1}{i} \sumf Q_f^2N\{\bar{f}(x)\gamma_\mu S^f(x-y)\gamma_\nu f(y)\no
\al\al\hspace{6em}+\bar{f}(y)\gamma_\nu S^f(y-x)\gamma_\mu f(x)\}\no
\al+\al Nj_\mu(x)j_\nu(y)\,,\eea
where $S^f(z)$ is the quark propagator and $N$ stands for normal ordering. In this expression, the singularities exclusively reside in the propagator -- the matrix elements of the normal ordered products are regular at $x=y$. The short distance expansion of the propagator starts with 
\be\label{eq:Sf}S^f(z)=-\frac{\gamma_\alpha z^\alpha}{2\pi^2(-z^2+i\epsilon)^2}+\frac{im_f}{4\pi^2(-z^2+i\epsilon)}+O(z^{-1})\,.\ee
The leading singularity is contained in the first line of equation \eqref{eq:TN} and is proportional to $(-z^2+i\epsilon)^{-3}$ -- the matrix elements thereof represent the disconnected part of the amplitude. We are interested in the singularities of the connected part, i.e.~in the terms that contain one quark propagator and two quark fields. To analyze these, we set $x=X+\frac{1}{2}z$, $y=X-\frac{1}{2}z$ and expand in powers of $z$. The expansion of the connected part starts with
\bea\label{eq:Thatz}\al\al T j_\mu(x)j_\nu(y) =\frac{\epsilon_{\mu\nu\alpha\beta}z^\alpha}{\pi^2(-z^2+i\epsilon)^2}\sumf Q_f^2\,\bar{f}\gamma^\beta\gamma_5f \\
\al\al\hspace{1em}- \frac{z^\alpha z^\beta}{2\pi^2(-z^2+i\epsilon)^2 }\sumf Q_f^2\,(g_{\mu\alpha}O_{\nu\beta}^f+g_{\nu\alpha}O_{\mu\beta}^f-g_{\mu\nu}O_{\alpha\beta}^f)\no
\al\al\hspace{1em}+\frac{g_{\mu\nu}}{2\pi^2(-z^2+i\epsilon)}\sumf Q_f^2\,\,m_f\bar{f} f+O(z^{-1})\,,\nonumber\eea
where $O_{\alpha\beta}^f$ stands for
\be\label{eq:oab}O_{\alpha\beta}^f=i \bar{f}\gamma_\alpha\partial\hspace{-0.6em}\rule{0em}{1em}^{\leftrightarrow}\hspace{-0.3em}_\beta f\,.\ee
We have dropped the normal ordering prescription as well as the argument of the quark fields -- it is understood that the quark bilinears occurring here are to be normal ordered and evaluated at the point $X$. 

The operator $O^f_{\alpha\beta}$ contains components with spin 0, 1 and 2: 
\bea O^f_{\alpha\beta}\al=\al O^{f_0}_{\alpha\beta}+ O^{f_1}_{\alpha\beta}+ O^{f_2}_{\alpha\beta}\,\\
O^{f_0}_{\alpha\beta}\al=\al\mbox{$\frac{1}{4}$}g_{\alpha\beta}O_{\,\lambda}^{f\,\lambda}\,,\no
O^{f_1}_{\alpha\beta}\al=\al\mbox{$\frac{1}{2}$}(O^f_{\alpha\beta}- O^f_{\beta\alpha})\,,\no
O^{f_2}_{\alpha\beta}\al=\al\mbox{$\frac{1}{2}$}(O^f_{\alpha\beta}+ O^f_{\beta\alpha}-\mbox{$\frac{1}{2}$}g_{\alpha\beta}
O_{\,\lambda}^{f\,\lambda})\,.\nonumber\eea
Neither the axial vector $\bar{f}\gamma^\beta\gamma_5 f$ nor the spin 1 operator $O^{f_1}_{\alpha\beta}$ contribute to the spin average.
The equation of motion relates the spin 0 part to the scalar operator $\bar{f}f$,
\be O_{\,\lambda}^{f\,\lambda}=2m_f\bar{f}f\,,\ee 
and $O^{f_2}_{\alpha\beta}$ is what becomes of the spin 2 operator specified in equation \eqref{eq:operators} when $g$ is set equal to zero.
Dropping terms that do not contribute to the spin average of the connected part, we obtain
\bea\label{eq:A} T j_\mu(x)j_\nu(y)\al=\al - \frac{z^\alpha z^\beta}{2\pi^2(-z^2+i\epsilon)^2 }\times\\
\al\al\hspace{-1em}\times\sumf Q_f^2(g_{\mu\alpha}O_{\beta\nu}^{f_2}+g_{\nu\alpha}O_{\beta\mu}^{f_2}-g_{\mu\nu}O_{\alpha\beta}^{f_2})\no
\al\al\hspace{-1em}-\frac{2z_\mu z_\nu+g_{\mu\nu}z^2 }{4\pi^2(-z^2+i\epsilon)^2}\sumf Q_f^2 m_f\bar{f} f+O(z^{-1})\,,\nonumber\eea

We denote the Fourier transform with respect to $z$ at fixed $X=\frac{1}{2}(x+y)$ by
\be\label{eq:Tz} \tilde{T}_{\mu\nu}(q,X)=\text{\small $\frac{i}{2}$}\!\int\!\! d^4z\,e^{i q\cdot z} Tj_\mu(x) j_\nu(y) \,.\ee
The OPE determines the behaviour of this quantity at large momenta. With the formulae
\bea\al\al\frac{i}{4\pi^2}\int \!\!d^4z e^{i q\cdot z} \frac{1}{-z^2+i\epsilon}=\frac{1}{-q^2-i\epsilon}\\
\al\al \frac{i}{4\pi^2}\int d^4z e^{i q\cdot z} \frac{z^\alpha z^\beta}{(-z^2+i\epsilon)^2}=\frac{-q^\alpha q^\beta+\frac{1}{2}g^{\alpha\beta}q^2}{(-q^2-i\epsilon)^2}\,,\nonumber\eea
the various terms occurring in \eqref{eq:A} yield:
\bea\al\al \tilde{T}_{\mu\nu}(q,X)=\no
\al\al\hspace{1em}\frac{q^\alpha q^\beta-\frac{1}{2}g^{\alpha\beta}q^2}{(-q^2-i\epsilon)^2}\sumf Q_f^2
 (g_{\mu\alpha}O_{\nu\beta}^{f_2}+g_{\nu\alpha}O_{\mu\beta}^{f_2}-g_{\mu\nu}O_{\alpha\beta}^{f_2})\no
\al\al\hspace{1em}+ \frac{q_\mu q_\nu-g_{\mu\nu}q^2}{(-q^2-i\epsilon)^2}\sumf Q_f^2m_f\bar{f}f +O(q^{-3})\,.\eea
The spin 2 part is indeed of the same structure as the coefficient of $c_2$ in equation \eqref{eq:Cc23}: for free quarks, the spin 2 coefficients are given by
\be\label{eq:c23PT}c_2^f(q^2)=\frac{Q_f^2}{(-q^2-i\epsilon)^2}\,,\hspace{1em}c_3^f(q^2)=0\,.\ee
Finally, comparison of the term proportional to $O^{f_0}=m_f\fbar f$ with equation \eqref{eq:Cf0} shows that for free quarks, the coefficient $c_1^f$ is given by
\be \label{eq:c1PT}c_1^f(q^2)=\frac{Q_f^2}{(-q^2-i\epsilon)^2}\,.\ee

\section{Derivation of the sum rule}
\label{sec:Derivation}
\setcounter{equation}{0}
To calculate the limiting value of the dispersion integral \eqref{eq:Tbardisp}, we first use partial fractions and split the integral into two parts:
\bea\label{eq:Tbardisp1}
\al\al \bar{T}^{\disp}=-\bar{S}^a+\bar{T}^{\disp'}\,,\\
\al\al \bar{S}^a =- \frac{m^2}{Q^2}\!\int_0^{\xth}\hspace{-1.3em}dx\,\frac{\bar{F}(x,Q^2)}{Q^2+m^2 x^2}\,,\no
\al\al \bar{T}^{\disp'} = \frac{4m^2\nu^2}{Q^2}\!\int_0^{\xth}\hspace{-1.3em}dx\,\frac{\bar{F}(x,Q^2)}{Q^4-4m^2 x^2\nu^2-i\epsilon}\,.\nonumber\eea
The reason for denoting the first term by $-\bar{S}^a$ is that it is independent of $\nu$ and hence amounts to a contribution to the subtraction 
function.  

To perform the limit in the remainder, we decompose the structure function into two parts with $\bar{F}=(\bar{F}-F^{\indR})+F^{\indR}$. In the difference $\bar{F}-\bar{F}^{\indR}$, the terms proportional to $x^{1-\alpha}$ cancel. We assume that the remainder is sufficiently smooth for $x\rightarrow 0$, so that $(\bar{F}-F^{\indR})/x^2$ is integrable and the integration can be interchanged with the limit.\footnote{If the singularity is more complicated, the sum rule does not get lost, but the explicit form must be adapted.} This leads to  
 \bea \al\al\bar{T}^{\disp'}\rightarrow -  \bar{S}^b +\bar{T}^{\disp''}\,, \\
 \al\al \bar{S}^b=\frac{1}{Q^2}\!\int_0^{\xth}\hspace{-1.3em}dx\,\frac{\bar{F}-\bar{F}^{\indR}}{x^2}\,, \no
 \al\al \bar{T}^{\disp''}=\sum_{\alpha>0}\frac{b_\alpha(Q^2)}{Q^2}J_\alpha\,,
  \hspace{1em}J_\alpha= \int_0^{\xth}\hspace{-1.3em}dx\,\frac{x^{1-\alpha}}{\xi^2-x^2-i\epsilon}\,,\nonumber\eea
with $\xi=Q^2/(2m \nu)$. The integral $J_\alpha$ represents a hypergeometric function.  What remains to be done is to work out the behaviour of this function at small values of  $\xi$. 

This can be done by making use of the known properties of the hypergeometric functions. Alternatively, one may observe that the contributions from the critical region $x\sim\xi$ remain the same if the integral is extended to infinity -- it can then be done explicitly.  On the interval $\xth < x < \infty$, on the other hand, the limit can be interchanged with the integral, which can then be done explicitly as well. The result reads
\be\label{eq:Jas} J_\alpha=-\frac{\pi}{2\sin\frac{1}{2}\pi \alpha} e^{-\frac{1}{2}i\pi\alpha} \xi^{-\alpha}+\frac{1}{\alpha\hspace{0.05em}\xth^\alpha}+O(\xi^2)\,.\ee
The first term is proportional to $\nu^\alpha$. The Reggeon amplitudes specified in equation \eqref{eq:TR} have the same behaviour when $\nu$ becomes large. Indeed, one readily checks that the two expressions agree, so that
\bea\label{eq:Tdisp2}\al\al\text{lim}\rule[-0.7em]{0em}{1em}_{\hspace{-1.7em}\nu\rightarrow\infty} (\bar{T}^{\disp''}-\bar{T}^{\indR})=- \bar{S}^c\\
\al\al\bar{S}^c=-\sum_{\alpha>0}\frac{b_\alpha(Q^2)}{\alpha\hspace{0.05em}\xth^\alpha}\,.\nonumber\eea  
Collecting terms, the formula \eqref{eq:Sbarlim} yields $
\bar{S}=\bar{S}^a+\bar{S}^b+\bar{S}^c$. This agrees with the expression \eqref{eq:Sum rule} for $\bar{S}$ quoted in section \ref{sec:Sum rule}. 

\end{document}